%% file: scalars.tex
\documentclass[preprint, tightenlines, a4paper, prd, twoside, showpacs, showkeys, superscriptaddress, nofootinbib]{revtex4}

\usepackage{amsmath}
\usepackage{amssymb}
\usepackage{bm}
\usepackage{graphicx}
\usepackage{latexsym}
\usepackage{mathrsfs}

\input{scalarsInput}

\begin{document}

\title{Comparison of Boltzmann Equations with\\
Quantum Dynamics for Scalar Fields}

\author{Manfred Lindner}
\email{lindner@ph.tum.de}
\affiliation{\TUM}

\author{Markus Michael M\"uller}
\email{Markus.Michael.Mueller@ph.tum.de}
\affiliation{\TUM}
\affiliation{\MPPMU}

\date{June 7, 2006}
\pacs{11.10.Wx, 98.80.Cq, 12.38.Mh}
\keywords{Nonequilibrium Quantum Field Theory, Kadanoff-Baym equations, 
          Boltzmann equations}
\preprint{TUM-HEP-613/05}
\preprint{MPP-2005-96}
\preprint{Phys.~Rev.~{\bf D73} (2006) 125002}

\begin{abstract}
Boltzmann equations are often used to study the thermal evolution of 
particle reaction networks. Prominent examples are the computation of the 
baryon asymmetry of the universe and the evolution of the quark-gluon plasma 
after relativistic heavy ion collisions. However, Boltzmann equations are 
only a classical approximation of the quantum thermalization process which is 
described by the so-called Kadanoff-Baym equations. This raises the question 
how reliable Boltzmann equations are as approximations to the full 
Kadanoff-Baym equations. Therefore, we present in this paper a detailed 
comparison between the Kadanoff-Baym and Boltzmann equations in the framework 
of a scalar $\Phi^4$ quantum field theory in 3+1 space-time dimensions.
The obtained numerical solutions reveal significant discrepancies in the 
results predicted by both types of equations. Apart from quantitative 
discrepancies, on a qualitative level the universality respected by the 
Kadanoff-Baym equations is severely restricted in the case of Boltzmann 
equations. Furthermore, the Kadanoff-Baym equations strongly separate the
time scales between kinetic and chemical equilibration. This separation of
time scales is absent for the Boltzmann equation.
\end{abstract}

\maketitle

\section{Introduction}

One of the most attractive frameworks to explain the matter-antimatter 
asymmetry of the universe is the so-called leptogenesis 
mechanism~\cite{Fukugita:1986hr,Buchmuller:2004nz,Buchmuller:2000nd}. 
Here, lepton number violating interactions in the
early universe produce a lepton asymmetry which is subsequently converted to 
the observed baryon asymmetry by so-called sphaleron processes. For the 
dynamical generation of the lepton asymmetry it is necessary, that the 
universe was in a state out of thermal equilibrium~\cite{Sakharov:1967dj}. 
The standard means to deal with this nonequilibrium situation are Boltzmann 
equations. However, it is well known that (classical) Boltzmann equations 
suffer from several shortcomings as compared to their quantum mechanical 
generalizations, the so-called Kadanoff-Baym equations. This motivates a 
comparison of Boltzmann and Kadanoff-Baym equations in order to assess the 
reliability of quantitative predictions of leptogenesis scenarios.

In addition to leptogenesis, there are various other systems which
warrant a comparison between Boltzmann and Kadanoff-Baym equations: 
In particular, a strong motivation is furnished by relativistic heavy 
ion collision experiments which aim at testing the quark-gluon plasma. 
In these experiments the quark-gluon plasma is produced in a state far 
from equilibrium. Recently, however, experiments claimed that the 
approach to thermal equilibrium should happen very fast, and that the 
evolution of the quark-gluon plasma could even be described by 
hydrodynamic 
equations~\cite{Arsene:2004fa,Back:2004je,Adams:2005dq,Adcox:2004mh}, 
which arise as approximations to Boltzmann equations. In this context it is
important to note that different quantities effectively thermalize on 
different time scales~\cite{Berges:2004ce}. Thus, one might face the 
situation that, although the full approach to thermal equilibrium takes a 
very long time, certain quantities, which are sufficient to describe the 
quark-gluon plasma with hydrodynamic equations, approach equilibrium values 
on much shorter time scales.

In order to derive Boltzmann equations from Kadanoff-Baym 
equations\footnote{The connection between Boltzmann equations and classical 
field theory has been treated in Refs.~\cite{Mueller:2002gd,Jeon:2004dh}.}, 
one has to employ several approximations, among them a first-order gradient
expansion, a Wigner transformation and a quasi-particle (or on-shell) 
approximation~\cite{baymKadanoff1962a,Danielewicz:1982kk,Ivanov:1999tj,
Knoll:2001jx,Blaizot:2001nr}. 
However, it is known, that the gradient expansion cannot be justified for 
early times. Consequently, one might expect that Boltzmann equations fail 
to describe the early-time evolution and that errors accumulated for early 
times cannot be remedied at late times. Of course, a Wigner transformation 
itself is not at all an approximation, but in order to make it available, one 
has to send the initial time to the remote past. Whereas Boltzmann equations 
imply the assumption of molecular chaos, meaning that two particles were 
uncorrelated before their collision, Kadanoff-Baym equations take these memory 
effects into account. Numerical solutions of Kadanoff-Baym equations revealed 
that this memory is lost gradually. Consequently, for late times it is indeed 
justifiable to send the initial time to the remote past. However, for early 
times this is certainly not the case. Furthermore, as a consequence of the
quasi-particle approximation, the conservation of momentum and energy prevents
Boltzmann equations from describing thermalization in $1+1$ space-time 
dimensions. In contrast to this, it has been shown  in the framework of a 
scalar $\Phi^4$ quantum field theory that this is feasible with Kadanoff-Baym 
equations~\cite{Berges:2001fi}. The reason for this qualitative discrepancy 
is that Kadanoff-Baym equations take off-shell effects into account 
\cite{Aarts:2001qa}, which are neglected in Boltzmann equations. 
Of course, in 3+1 dimensions both types of equations are capable of describing
thermalization. In the case of leptogenesis, however, the on-shell character 
of the Boltzmann equation leads to a further inconsistency: All leptogenesis 
scenarios share the fact that some heavy particles decay out of thermal 
equilibrium into the particles which we observe in the universe today. The 
spectral function of a particle that can decay into other particles is given 
by a Breit-Wigner curve with a non-vanishing width. By employing the 
quasi-particle approximation we reduce this decay width of the particles to 
zero, i.e.~a Boltzmann equation can only describe systems consisting of 
stable, or at least very long-lived, particles. After all, how does the 
on-shell character of the Boltzmann equation affect the description of 
quantum fields out of thermal equilibrium in $3+1$ dimensions?

When applying Boltzmann equations to the description of leptogenesis,
the standard technique to construct the collision integrals --- before 
employing further approximations --- is to take the usual bosonic and 
fermionic statistical gain and loss terms multiplied with the S-matrix element
for the respective reaction \cite{Kolb:1979qa,kolbTurner1990a}. These 
S-matrix elements are computed in vacuum, and one may wonder of which 
significance they are for a statistical quantum mechanical system.

All these shortcomings of Boltzmann equations lead to the conclusion that 
one should perform a detailed comparison between Boltzmann and Kadanoff-Baym 
equations \cite{Danielewicz:1982kk,kohler1995a,Morawetz:1998em,
Juchem:2003bi}, such that one can explicitely see how large the 
quantum mechanical corrections are. Due to the complexity of the problem, we 
restrict ourselves for the moment to a $\Phi^4$ quantum field theory in $3+1$ 
space-time dimensions. Of course, in this framework one can neither describe 
the phenomenon of leptogenesis nor thermalization after a heavy ion collision.
Nevertheless, it may well serve as starting point for further research, and
certainly permits to present a detailed comparison between Boltzmann and 
Kadanoff-Baym equations, which may point to interesting phenomena to be 
investigated in more realistic theories.

In general, when studying systems out of thermal equilibrium by means of
Kadanoff-Baym equations, it is crucial to start from a $\Phi$-derivable 
approximation, since these approximations ensure the conservation of energy
and global charges~\cite{baymKadanoff1961a,Baym:1962sx,Ivanov:1998nv}. The 2PI 
effective action furnishes such a $\Phi$-derivable 
approximation~\cite{Jackiw:1974cv,Cornwall:1974vz,Calzetta:1986cq} and has 
proven to be an efficient and reliable tool for
the description of out-of-equilibrium quantum fields in numerous previous 
treatments~\cite{Berges:2000ur,Berges:2001fi,Berges:2002wr,Aarts:2001yn,
Aarts:2003bk,Berges:2004hn}. In this work, we start from the 2PI 
effective action truncated at three-loop order. The Kadanoff-Baym equations 
can be obtained by requiring that the 2PI effective action be stationary with 
respect to variations of the full connected two-point 
function~\cite{Aarts:2001qa,Berges:2001fi}. In order to derive the 
corresponding Boltzmann equation, subsequently one has to employ a 
gradient expansion, a Wigner transformation, the Kadanoff-Baym ansatz and the 
quasi-particle approximation~\cite{baymKadanoff1962a,Danielewicz:1982kk,
Ivanov:1999tj,Knoll:2001jx,Blaizot:2001nr}. While the Boltzmann equation 
describes the time evolution of the particle number distribution, the 
Kadanoff-Baym equations describe the evolution of the full quantum mechanical 
two-point function of the system. However, one can define an effective 
particle number distribution which is given by the full propagator and its 
time derivatives evaluated at equal times~\cite{Berges:2001fi,Berges:2002wr}. 
Finally, we solve the Boltzmann and the Kadanoff-Baym equations numerically 
for spatially homogeneous and isotropic systems in 3+1 dimensions and compare 
their predictions on the evolution of these systems for various initial 
conditions.

\section{2PI Effective Action}

\begin{figure*}[tb]
  \centering
  \includegraphics{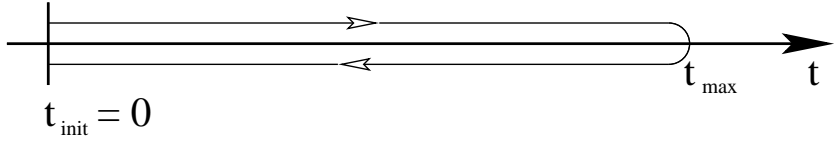}
  \caption{\label{fig2.1} Closed real-time path $\c$. This time path 
  was invented by Schwinger~\cite{Schwinger:1960qe} (see 
  also~\cite{Bakshi:1962dv,Bakshi:1963bn}) and applied to nonequilibrium 
  problems by Keldysh~\cite{Keldysh:1964ud}. In order to avoid the doubling 
  of the degrees of freedom, we use the form presented in 
  Ref.~\cite{Danielewicz:1982kk}.}
\end{figure*}

In this work we consider a real scalar quantum field, whose dynamics is 
determined by the Lagrangian density
\[ \mathscr{L} = - \frac{1}{2} \Big( \partial_{\mu} \Phi \Big) 
\Big( \partial^{\mu} \Phi \Big) - \frac{1}{2} m_B^2 \Phi^2 - \frac{\lambda}{4 !} \Phi^4 \;. \]
The minus sign of the kinetic term indicates that we use the metric where the
time-time component is negative. As we will compute the evolution of the 
two-point Green's function for a nonequilibrium many body system, already 
the classical action has to be defined on the closed Schwinger-Keldysh 
real-time contour, shown in Fig.~\ref{fig2.1}. 
The free inverse propagator can then be read off the free part of the 
classical action
\[ I_0 = - \frac{1}{2} \intl_{\c} \dddd{x} \intl_{\c} \dddd{y} 
\left[ \Phi \left( x \right) G_0^{-1} \left( x, y \right) \Phi \left( y \right) \right] \;, \]
where
\begin{equation} \label{eq2.1}
  G_0^{-1} \left( x, y \right) = \left( \partial_{x^{\mu}} \partial_{y_{\mu}}
  + m_B^2 \right) \delta_{\c} \left( x - y \right) \;.
\end{equation}
We consider a system without symmetry breaking, i.e. $\left\langle \Phi 
\left( x \right) \right\rangle = 0$. In this case the full connected 
Schwinger-Keldysh propagator is given by
\[ G \left( x, y \right) = \left\langle T_{\c} \left\{ \Phi \left( x \right) 
\Phi \left( y \right) \right\} \right\rangle \;. \]
Accordingly, for Gaussian initial conditions the 2PI effective action can be 
parameterized in the form~\cite{Jackiw:1974cv,Cornwall:1974vz,
Calzetta:1986cq,Berges:2001fi}
\[ \Gamma \left[ G \right] = \frac{i}{2} \tr_{\c} \log_{\c} \left[ G^{-1} \right] - \frac{1}{2} \tr_{\c} \left[ G_0^{-1} G \right] + \Gamma_2 \left[ G \right] + const \;. \]
$i \Gamma_2 \left[ G \right]$ is the sum of all two-particle irreducible
vacuum diagrams, where internal lines represent the full connected propagator
$G \left( x, y \right)$. Of course, for an interacting theory we cannot 
compute $\Gamma_2 \left[ G 
\right]$ completely, and we have to rely on approximations. In this work we 
apply the loop expansion of the 2PI effective action up to three-loop order.
The diagrams contributing to $\Gamma_2 \left[ G \right]$ in this 
approximation are shown in Fig.~\ref{fig2.2}. We find~\cite{Aarts:2001qa}:
\begin{eqnarray*}
        \Gamma_2 \left[ G \right] 
  & = & - \frac{\lambda}{8} \intl_{\c} \dddd{x} \left[ G \left( x, x \right) G \left( x, x \right) \right] \\
  &   & {} + \frac{i \lambda^2}{48} \intl_{\c} \dddd{x} \intl_{\c} \dddd{y} \left[ G \left( x, y \right) G \left( x, y \right) G \left( y, x \right) G \left( y, x \right) \right] \;.
\end{eqnarray*}

\section{Kadanoff-Baym Equations}

The equation of motion for the full propagator reads 
\cite{Jackiw:1974cv,Cornwall:1974vz}
\[ \frac{\delta \Gamma \left[ G \right]}{\delta G \left( y, x \right)} = 0 \;. \]
It is equivalent to the Schwinger-Dyson equation
\begin{equation} \label{eq2.2}
  G^{-1} \left( x, y \right) = i G^{-1}_0 \left( x, y \right) - \Pi \left( x, y \right) \;,
\end{equation}
where the proper self energy is given by
\begin{equation}
 \Pi \left( x, y \right) 
   =  2 i \frac{\delta \Gamma_2 \left[ G \right]}{\delta G \left( y, x \right)} 
   =  - \frac{i \lambda}{2} \delta_{\c} \left( x - y \right) G \left( x, x \right) - \frac{\lambda^2}{6} G \left( x, y \right) G \left( x, y \right) G \left( y, x \right) \label{eq2.3} \;.
\end{equation}
After we have inserted the free inverse propagator given in Eq.~(\ref{eq2.1}),
we convolute the Schwinger-Dyson equation (\ref{eq2.2}) with $G$ from the right:
\begin{figure*}
  \centering
  \includegraphics{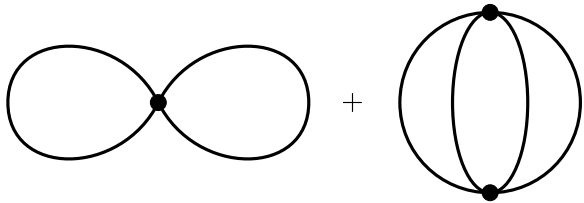}
  \caption{\label{fig2.2}Two- and three-loop contribution to $\Gamma_2 
           \left[ G \right]$. The lines represent the full connected
           Schwinger-Keldysh propagator.}
\end{figure*}
\begin{equation} \label{eq2.4}
  i \left( - \partial_{x^{\mu}} \partial_{x_{\mu}} + m_B^2 \right) G \left( x, y \right) = \delta_{\c} \left( x - y \right) + \intl_{\c} \dddd{z} \left[ \Pi \left( x, z \right) G \left( z, y \right) \right]
\end{equation}
Next, we define the spectral function\footnote{From the definition of the 
spectral function we see that it is antisymmetric in the sense that 
$G_{\varrho} \left( x, y \right) = - G_{\varrho} \left( y, x \right)$. 
Furthermore, the canonical equal-time commutation relations give 
$\left( G_{\varrho} \left( x, y \right) \right)_{x^0 = y^0} = 0$ and 
$\left( \partial_{y^0} G_{\varrho} \left( x, y \right) \right)_{x^0 = y^0}
= - \delta^3 \left( \bm{x} - \bm{y} \right)$.}
\[ G_{\varrho} \left( x, y \right) = i \left\langle \left[ \Phi \left( x \right), \Phi \left( y \right) \right]_- \right\rangle \]
and the statistical propagator\footnote{In contrast to the spectral function, 
the statistical propagator is symmetric in the sense that 
$G_F \left( x, y \right) = G_F \left( y, x \right)$.}
\[ G_F \left( x, y \right) = \frac{1}{2} \left\langle \left[ \Phi \left( x \right), \Phi \left( y \right) \right]_+ \right\rangle \]
such that we can write the full propagator as
\begin{equation} \label{eq2.5}
  G \left( x, y \right) = G_F \left( x, y \right) - \frac{i}{2} \sign_{\c} \left( x^0 - y^0 \right) G_{\varrho} \left( x, y \right) \;. 
\end{equation}
Note that for real scalar quantum fields both the statistical propagator and
the spectral function are real-valued functions \cite{Berges:2001fi}. The 
spectral function describes the particle spectrum of our theory. From its 
Wigner transform we 
can obtain the thermal mass and the decay width of the particles in our 
system. On the other hand we will define an effective particle number density 
given by the statistical propagator and its time derivatives evaluated at 
equal times. From Eq.~(\ref{eq2.3}) (and Fig.~\ref{fig2.3}) 
we see that the self energy contains a local and a nonlocal part:
\[ \Pi \left( x, y \right) = - i \delta_{\c} \left( x - y \right) \Pi^{(local)} \left( x \right) + \Pi^{(nonlocal)} \left( x, y \right) \;. \]
The local part of the self energy only causes a mass shift, which can be 
included in an effective mass:
\begin{equation} \label{eq2.7}
  M^2 \left( x \right) = m_B^2 + \Pi^{(local)} \left( x \right) = m_B^2 + \frac{\lambda}{2} G_F \left( x, x \right) \;.
\end{equation}
After inserting Eq.~(\ref{eq2.5}) into Eq.~(\ref{eq2.3}), we can decompose the 
nonlocal part of the self energy in exactly the same way as we did for the 
propagator:
\begin{figure*}[tb]
  \centering
  \includegraphics{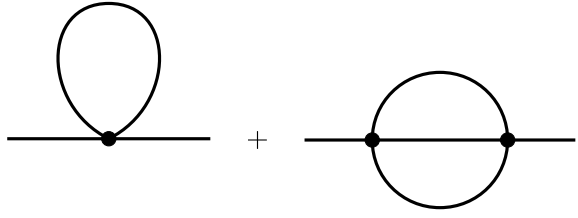}
  \caption{\label{fig2.3}One- and two-loop contribution to the proper 
           self-energy $\Pi$. Again, internal lines represent the full
           connected Schwinger-Keldysh propagator. The tadpole represents 
           the local part which causes a mass shift only. The setting-sun 
           diagram represents the nonlocal part and leads to 
           thermalization.}
\end{figure*}
\[ \Pi^{(nonlocal)} \left( x, y \right) = \Pi_F \left( x, y \right) - \frac{i}{2} \sign_{\c} \left( x^0 - y^0 \right) \Pi_{\varrho} \left( x, y \right) \;. \]
We find
\[ \Pi_F \left( x, y \right) = - \frac{\lambda^2}{6} \Big( G_F \left( x, y \right) G_F \left( x, y \right) G_F \left( x, y \right) - \frac{3}{4} G_{\varrho} \left( x, y \right) G_{\varrho} \left( x, y \right) G_F \left( x, y \right) \Big) \]
and
\[ \Pi_{\varrho} \left( x, y \right) = - \frac{\lambda^2}{6} \Big( 3 G_F \left( x, y \right) G_F \left( x, y \right) G_{\varrho} \left( x, y \right) - \frac{1}{4} G_{\varrho} \left( x, y \right) G_{\varrho} \left( x, y \right) G_{\varrho} \left( x, y \right) \Big) \;. \]
When we insert all these definitions into Eq.~(\ref{eq2.4}), we observe that it
splits into two complementary real-valued evolution equations for the 
statistical propagator and the spectral function, 
respectively~\cite{Berges:2001fi}. These are the so-called Kadanoff-Baym 
equations:
\begin{equation} \label{eq2.6}
\left( - \partial_{x^{\mu}} \partial_{x_{\mu}} + M^2 \left( x \right) \right) G_F \left( x, y \right)
   =  \intl_{0}^{y^0} \dddd{z} \; \Pi_F \left( x, z \right) G_{\varrho} \left( z, y \right) 
      - \intl_{0}^{x^0} \dddd{z} \; \Pi_{\varrho} \left( x, z \right) G_F \left( z, y \right)
\end{equation}
and
\begin{equation} \label{eq2.12}
  \left( - \partial_{x^{\mu}} \partial_{x_{\mu}} + M^2 \left( x \right) \right) G_{\varrho} \left( x, y \right) = - \intl_{y^0}^{x^0} \dddd{z} \; \Pi_{\varrho} \left( x, z \right) G_{\varrho} \left( z, y \right) \;.
\end{equation}
For a spatially homogeneous system, one can Fourier transform these equations 
with respect to the spatial relative coordinate. Furthermore, in an isotropic
system the propagator will depend only on the modulus of the momentum. 
As explained in more detail in Refs.~\cite{Berges:2001fi,Berges:2002wr}, one 
can define effective kinetic energy and particle number densities $\omega 
\left( t, \bm{p} \right)$ and $n \left( t, \bm{p} \right)$ which are given by 
\begin{equation} \label{eq2.22}
  \omega^2 \left( t, \bm{p} \right) = \left( \frac{\partial_{x^0} \partial_{y^0} G_F \left( x^0, y^0, \bm{p} \right)}{G_F \left( x^0, y^0, \bm{p} \right)} \right)_{x^0 = y^0 = t}
\end{equation}
and
\begin{equation} \label{eq2.23}
  n \left( t, \bm{p} \right) = \omega \left( t, \bm{p} \right) G_F \left( t, t, \bm{p} \right) - \frac{1}{2} \;.
\end{equation}
However, we stress that the Kadanoff-Baym equations are self-consistent 
evolution equations for the full propagator of our system, and that one has
to follow the evolution of the two-point function throughout the whole
$x^0$-$y^0$-plane (of course, constrained to the part with $x^0 \ge 0$ and 
$y^0 \ge 0$). One can then follow the evolution of the effective particle
number density along the bisecting line of this plane.

We would like to emphasize that the only approximation involved in the 
numerical solution of the Kadanoff-Baym equations is the loop expansion of
the 2PI effective action. In the next section we will describe the 
approximations which are necessary to derive a Boltzmann equation from the
Kadanoff-Baym equation for the statistical propagator (\ref{eq2.6}).

\section{Boltzmann Equations}

It is well known how Boltzmann equations can be obtained as an approximation 
of the Kadanoff-Baym 
equations~\cite{baymKadanoff1962a,Danielewicz:1982kk,Blaizot:2001nr}. 
In this section we briefly review the standard derivation: 
One has to employ a Wigner transformation, a gradient expansion, 
the Kadanoff-Baym ansatz and the quasi-particle approximation.

\begin{figure*}[tb]
  \hspace*{\fill}
  \includegraphics[width=80mm]{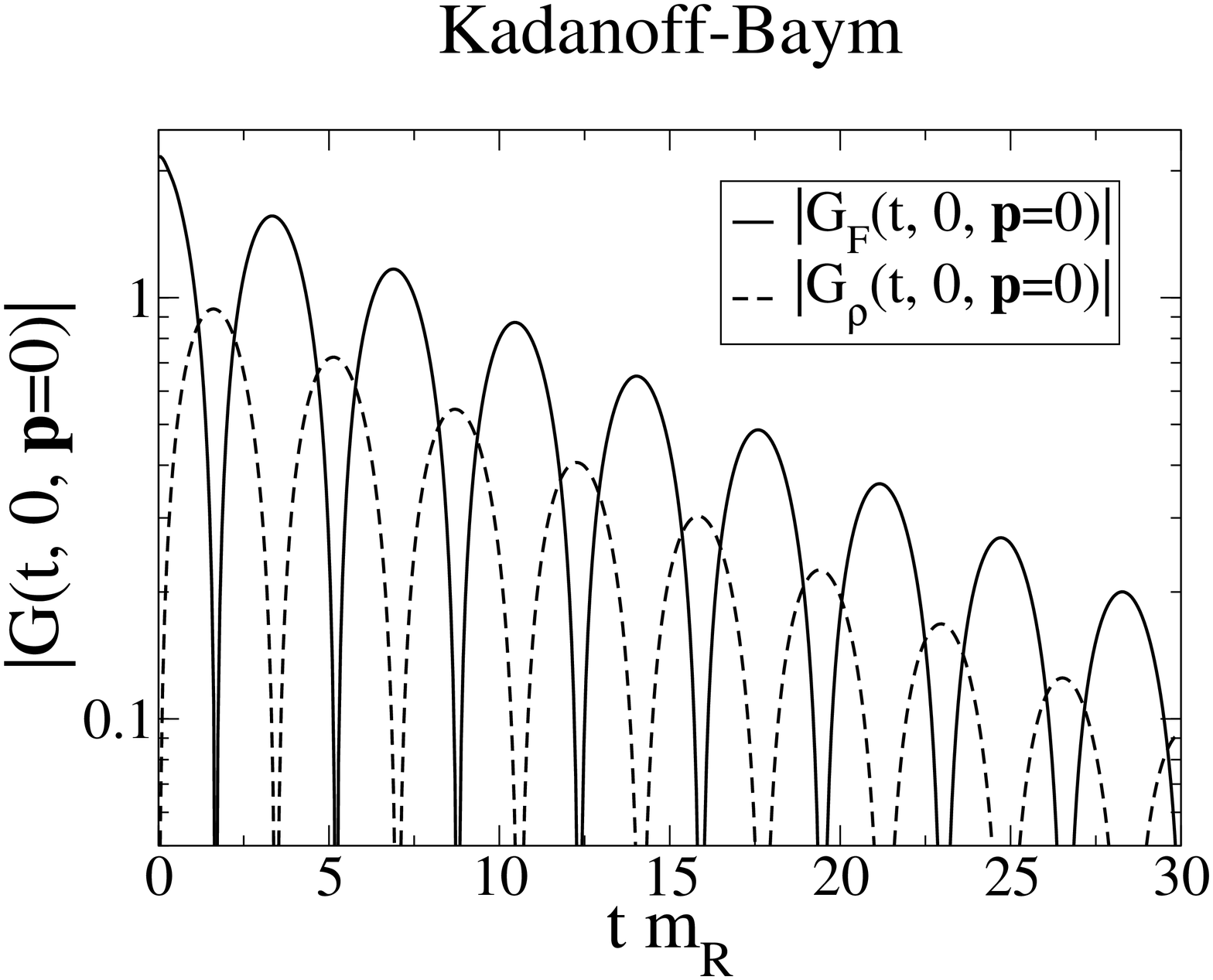}
  \hfill \hfill
  \includegraphics[width=80mm]{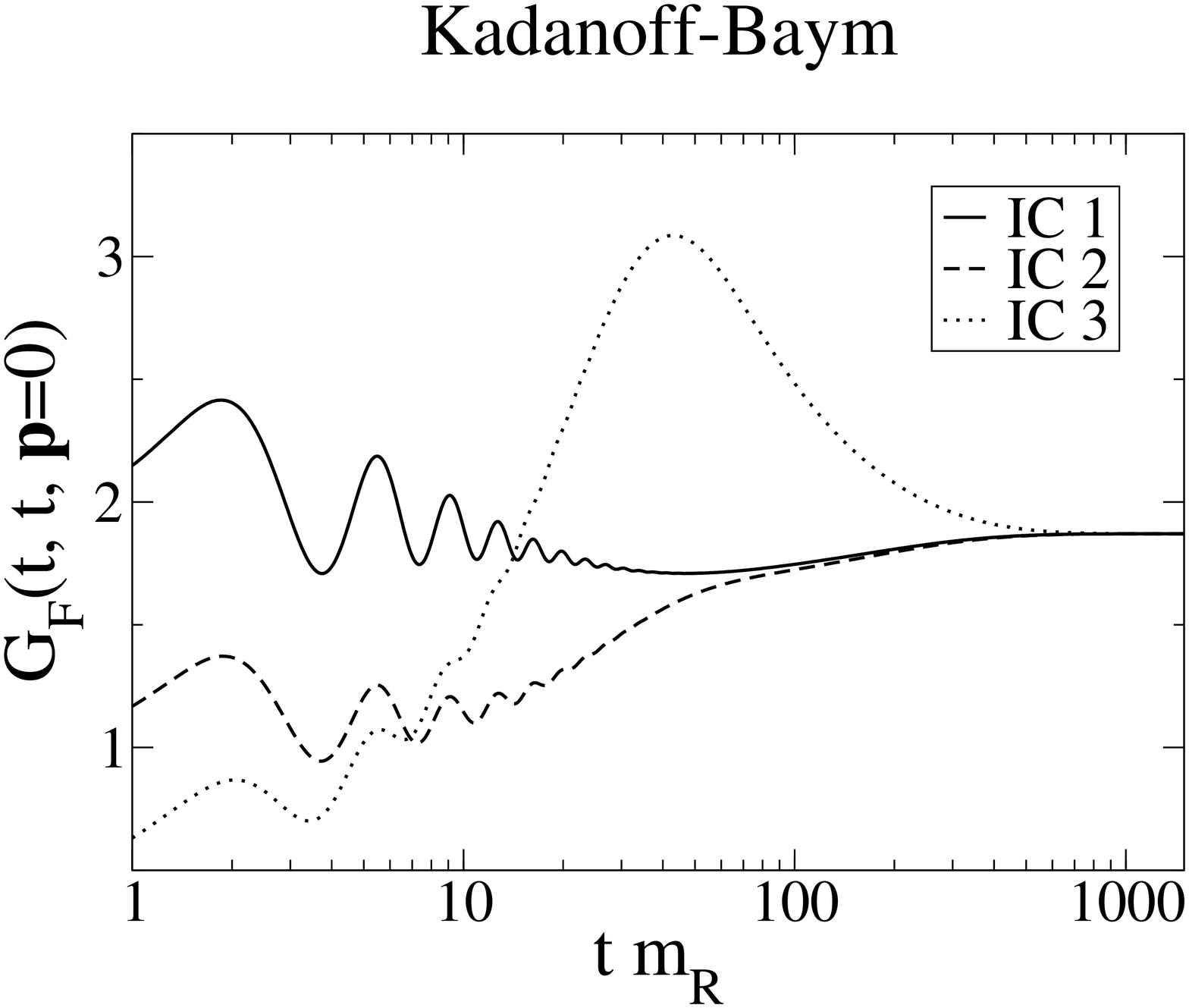}
  \hfill \\
  \hspace*{\fill}
  \begin{minipage}[t]{75mm}
    \caption{\label{fig2.5}
             The modulus of the unequal-time propagator as function of time 
             for fixed momentum mode $\bm{p} = 0$. Correlations between 
             earlier and later times are exponentially damped.}
  \end{minipage}
  \hfill \hfill
  \begin{minipage}[t]{75mm}
    \caption{\label{fig2.4}The equal-time propagator as a function of time 
             for three different initial conditions (cf.~Fig.~\ref{fig2.6}). 
             The system shows rapid oscillations which die out after moderate 
             times and are followed by a smooth drifting regime.}
  \end{minipage} 
  \hfill
\end{figure*}

First, we subtract the Hermitian adjoint of Eq.~(\ref{eq2.6}) from 
Eq.~(\ref{eq2.6}) and re-parameterize the propagator and the self energy by
center and relative coordinates
\[ G \left( u, v \right) = \tilde{G} \left( \frac{u+v}{2}, u-v \right) \;. \]
Next, we define $X = \frac{x+y}{2}$ and $s = x - y$, and observe on the left
hand side of the difference equation that
\[ - \partial_{x^{\mu}} \partial_{x_{\mu}} + \partial_{y^{\mu}} \partial_{y_{\mu}} = - 2 \partial_{X^{\mu}} \partial_{s_{\mu}} \]
is automatically of first order in $\partial_X$. Furthermore, we Taylor expand
the effective masses on the left hand side as well as the propagators and self
energies on the right hand side to first order in $\partial_X$ around $X$. 
After that, we Fourier transform the difference equation with respect to $s$. 
The Wigner transformed statistical propagator and spectral function are given 
by
\[ \tilde{G}_F \left( X, k \right) = \int \dddd{s} \; \exp \left( - iks \right) \tilde{G}_F \left( X, s \right) \]
and
\[ \tilde{G}_{\varrho} \left( X, k \right) = - i \int \dddd{s} \; \exp \left( - iks \right) \tilde{G}_{\varrho} \left( X, s \right) \;. \]
The factor of $- i$ in the Wigner transform of the spectral 
function makes $\tilde{G}_{\varrho} \left( X, k \right)$ again a 
real-valued function. However, in order to be able to really perform the 
Fourier transformation, we have to send the initial time to $- \infty$. At 
least for large $x^0$ and $y^0$ this can be justified by taking into account 
that correlations between earlier and later times are suppressed 
exponentially, as one can see in Fig.~\ref{fig2.5}. 
The result of all these transformations is 
a quantum kinetic equation for the statistical propagator\footnote{The 
retarded propagator $G_R \left( x, y \right) = \theta \left( x^0 - y^0 \right) 
G_{\varrho} \left( x, y \right)$ and self energy, as well as the 
corresponding advanced quantities, have to be introduced in order to remove 
the upper boundaries of the memory integrals. As a result the complete 
system of quantum kinetic equations includes six equations: one equation
for $G_F$, $G_{\varrho}$, $G_R$, $\Pi_F$, $\Pi_{\varrho}$ and $\Pi_R$, 
respectively.}~\cite{Ivanov:1999tj,Knoll:2001jx,Blaizot:2001nr,Berges:2002wt,
Juchem:2004cs,Berges:2005vj,Berges:2005md}:
\begin{eqnarray}
        \lefteqn{\left( 2 k^{\mu} \partial_{X^{\mu}} - \left( \partial_{X^{\mu}} M^2 \left( X \right) \right) \partial_{k_{\mu}} \right) \tilde{G}_F \left( X, k \right)} \; \nonumber \\
  & = & \tilde{\Pi}_{\varrho} \left( X, k \right) \tilde{G}_F \left( X, k \right) - \tilde{\Pi}_F \left( X, k \right) \tilde{G}_{\varrho} \left( X, k \right) \label{eq2.21} \\
  &   & {} + \left\{ \tilde{\Pi}_F \left( X, k \right) ; \Re \left( \tilde{G}_R \left( X, k \right) \right) \right\}_{PB} + \left\{ \Re \left( \tilde{\Pi}_R \left( X, k \right) \right) ; \tilde{G}_F \left( X, k \right) \right\}_{PB} \nonumber \;,
\end{eqnarray}
where the Poisson brackets are defined by
\begin{equation}
  \left\{ \tilde{f} \left( X, k \right) ; \tilde{g} \left( X, k \right) \right\}_{PB}
  = \Big[ \partial_{X^{\mu}} \tilde{f} \left( X, k \right) \Big] \Big[ \partial_{k_{\mu}} \tilde{g} \left( X, k \right) \Big]
    - \Big[ \partial_{k_{\mu}} \tilde{f} \left( X, k \right) \Big] \Big[ \partial_{X^{\mu}} \tilde{g} \left( X, k \right) \Big] \;.
\end{equation}
Employing the first order Taylor expansion is clearly not justifiable for 
early times when the equal-time propagator is rapidly oscillating, 
cf.~Fig.~\ref{fig2.4}. But this 
is obvious, since employing this gradient expansion is clearly motivated by 
equilibrium considerations: In equilibrium the propagator depends on the 
relative coordinates only. There is no dependence on the center coordinates, 
and one may hope that there are situations where the propagator depends only 
moderately on the center coordinates. This is clearly the case for late times 
when our system is sufficiently close to equilibrium. However, as is shown in
Fig.~\ref{fig2.4}, already after moderate times the rapid oscillations 
mentioned above, have died out and are followed by a smooth drifting 
regime~\cite{Berges:2001fi}. In this drifting regime the second derivative 
with respect to $X$ should be negligible as compared to the first order 
derivative and a consistent Taylor expansion can be justified even though the 
system may still be far from equilibrium. However, it is crucial that the 
Taylor expansion is performed consistently for two reasons: First, this 
guarantees that the quantum kinetic equations satisfy exactly the same 
conservation laws as the full Kadanoff-Baym equations do~\cite{Knoll:2001jx}. 
Second, it has been shown that neglecting the Poisson brackets severely 
restricts the range of validity of the quantum kinetic 
equations~\cite{Berges:2005ai,Berges:2005md}. For the intermediate and 
late-time regimes these quantum kinetic equations have the advantage 
that they do not include any memory integrals. Being local in time, their 
numerical solution requires much less computer memory as compared to the 
Kadanoff-Baym equations and algorithms using an adaptively controlled 
time-step size become available. Furthermore, the energy convolutions 
replacing the memory integrals can be done quite efficiently using a Fast 
Fourier Transform algorithm. In order to derive a Boltzmann equation from the 
quantum kinetic equation (\ref{eq2.21}), first we have to discard the Poisson 
brackets, thereby sacrificing the consistency of the gradient expansion. After 
that, we employ the Kadanoff-Baym ansatz
\begin{equation} \label{eq2.10} 
  \tilde{G}_F \left( X, k \right) = \tilde{G}_{\varrho} \left( X, k \right) 
  \left( \tilde{n} \left( X, k \right) + \frac{1}{2} \right) \;,
\end{equation}
which also can be motivated by equilibrium considerations. In fact, this is a 
generalization of the fluctuation-dissipation theorem, which states that, for 
a system in thermal equilibrium, the statistical propagator is proportional to
the spectral function. The fluctuation dissipation theorem can be recovered 
from Eq.~(\ref{eq2.10}) by discarding the dependence on the center coordinate 
$X$ and fixing $\tilde{n}$ to be the Bose-Einstein distribution function. The 
last approximation, which is necessary to arrive at the Boltzmann equation, is
the so-called quasi-particle (or on-shell) approximation:
\begin{equation} \label{eq2.11} 
  \tilde{G}_{\varrho} \left( X, k \right) = \frac{\pi}{E \left( X, \bm{k} \right)} \Big( \delta \left( k^0 - E \left( X, \bm{k} \right) \right) - \delta \left( k^0 + E \left( X, \bm{k} \right) \right) \Big) \;,
\end{equation}
where the quasi-particle energy is given by
\[ E \left( X, \bm{k} \right) = \sqrt{m_{th}^2 + \bm{k}^2} \;. \]
Once more, we would like to stress that the exact time evolution of the 
spectral function is determined by the Kadanoff-Baym equations. It has been 
shown that the spectral function can be parameterized by a Breit-Wigner 
function with a non-vanishing width~\cite{Aarts:2001qa,Juchem:2003bi}. To 
reduce the width of this Breit-Wigner curve to zero is certainly not a 
controllable approximation and leads to significant qualitative discrepancies 
between the results produced by Kadanoff-Baym and Boltzmann equations. In fact 
this approximation can only be justified if our system consists of stable, or 
at least very long-lived, quasi-particles, whose mass is much larger than 
their decay width. We also would like to note that a completely 
self-consistent determination of the thermal mass in the framework of the 
Boltzmann equation requires the solution of an integral equation for
$E \left( X, \bm{k} \right)$, which would drastically increase the complexity
of our numerics. As none of our physical results depend on the exact value of
the thermal mass, for convenience, we set $m_{th}$ to the equilibrium value 
of the thermal mass as determined by the Kadanoff-Baym equations.
Eventually, we define the quasi-particle number density by
\[ n \left( X, \bm{k} \right) = \tilde{n} \left( X, \bm{k}, E \left( X, \bm{k} \right) \right) \;. \]
After equating the positive energy components in Eq.~(\ref{eq2.21}) we arrive 
at the Boltzmann equation. For a spatially homogeneous system there is no
dependence on the spatial center coordinates and the Boltzmann equation 
reads\footnote{Here, we use the abbreviations $E_k = \sqrt{m_{th}^2 + 
\bm{k}^2}$ and $n_{\bm{k}} = n \left( t, \bm{k}\right)$.}:
\begin{eqnarray}
        \lefteqn{\partial_t n \left( t, \bm{k} \right) = \frac{\lambda^2 \pi}{48} \int \frac{\ddd{p}}{\left( 2 \pi \right)^3} \int \frac{\ddd{q}}{\left( 2 \pi \right)^3} \int \ddd{r} \Bigg[ \frac{1}{E_k E_p E_q E_r}} \nonumber \\
  &   & {} \times \delta \left( \bm{k} + \bm{p} - \bm{q} - \bm{r} \right) \delta \left( E_k + E_p - E_q - E_r \right) \label{eq2.17} \\
  &   & {} \times \Big( \left( 1 + n_{\bm{k}} \right) \left( 1 + n_{\bm{p}} \right) n_{\bm{q}} n_{\bm{r}} - n_{\bm{k}} n_{\bm{p}} \left( 1 + n_{\bm{q}} \right) \left( 1 + n_{\bm{r}} \right) \Big) \Bigg] \;. \nonumber
\end{eqnarray}
For a spatially homogeneous and isotropic system we can dramatically simplify 
the collision integral~\cite{Dolgov:1997mb}, which allows us to reduce the
complexity of our numerics significantly. The details of this calculation are 
shown in the appendix. The result is the following equation\footnote{Now, 
$k = \left| \bm{k} \right|$ and $n_k = n \left( t, \left| \bm{k} \right| 
\right)$}:
\begin{eqnarray}
        \lefteqn{\partial_t n \left( t, k \right) = \frac{\lambda^2}{96 \pi^4} \intl_0^{\infty} dp \intl_0^{\infty} dq \Bigg[ \Theta \left( r_0^2 \right) \frac{p q D \left( k, p, q, r_0 \right)}{E_k E_p E_q}} \label{eq2.9} \\
  &   & {} \times \Big( \left( 1 + n_k \right) \left( 1 + n_p \right) n_q n_{r_0} - n_k n_p \left( 1 + n_q \right) \left( 1 + n_{r_0} \right) \Big) \Bigg] \;. \nonumber
\end{eqnarray}
The auxiliary functions $r_0$ and $D$ are obtained from very simple 
expressions which are given in the appendix.
In this section we have shown that, using a gradient expansion 
and a quasi-particle (or on-shell) approximation, one can derive the Boltzmann
equation from the Kadanoff-Baym equations. In this sense one can consider the
Kadanoff-Baym equations as quantum Boltzmann equations re-summing the
gradient expansion up to infinite order and including off-shell and memory 
effects. In the next section we are going to explain how we solved the 
Boltzmann and the Kadanoff-Baym equations numerically.

\section{Numerical Implementation}

\subsection{Kadanoff-Baym Equations}

For the numerical solution of the Kadanoff-Baym equations we follow exactly
the lines of Refs.~\cite{Berges:2001fi, Berges:2004yj, montvayMunster1994a}, 
i.e.~for the spatial coordinates we employ a standard 
discretization on a three-dimensional lattice with lattice spacing $a_s$ and 
$N_s$ lattice sites in each direction. Thus, the lattice momenta are given by
\[ \hat{p}_{n_j} = \frac{2}{a_s} \sin \left( \frac{\pi n_j}{N_s} \right) \;, \]
where $n_j$, $j \in \left\{ 1, 2, 3 \right\}$, enumerates the momentum modes 
in the $j$-th dimension. As we consider a spatially homogeneous and isotropic 
system, for given times $\left( x^0, y^0 \right)$ we only need to store the 
propagator for momentum modes with $\frac{N_s}{2} \ge n_1 \ge n_2 \ge n_3 \ge 
0$. This saves us a factor of 48 in memory usage. The discretization in time 
leads to a history matrix $H = \left\{ 0, a_t, 2 a_t, \ldots, \left( N_t - 1
\right) a_t \right\}^2$. Here $a_t$ is the step size and $N_t$ is the number 
of times in each time dimension for which we keep the propagator in memory in 
order to compute the memory integrals. This history cut off can be justified
by the exponential damping of the unequal-time propagator, 
cf.\ Fig.~\ref{fig2.5}. Exploiting the symmetry of the 
statistical propagator with respect to the interchange of its time arguments, 
we only need to store the values of the statistical propagator for $x^0 \ge 
y^0$. In very much the same way we can use the respective antisymmetry of the 
spectral function. This saves us another factor of 2 in memory usage. The 
convolutions arising in the computation of the setting-sun self-energies are 
most efficiently computed using a Fast Fourier Transform algorithm for 
real-valued even functions~\cite{fftw}.

In order to set the scale for the simulations, we use the renormalized vacuum 
mass $m_R$. The corresponding bare mass $m_B$ is obtained through a 
perturbative renormalization at one-loop order of the self energy 
(tadpole)~\cite{Arrizabalaga:2005tf}. We solved the Kadanoff-Baym 
equations numerically on a lattice with $N_t = 500$, $N_s = 32$, 
$a_t m_R = 0.06$, $a_s m_R = 0.5$ and $\lambda = 18$. So far we performed our 
simulations on a simple desktop PC with a Pentium4 processor and 2 GByte RAM. 
However, we would like to note that the numerics can easily be parallelized.

\subsection{Boltzmann Equation}

As we saw in the previous paragraph, in order to discretize the Kadanoff-Baym
equations we can rely on the well-defined scheme offered by standard lattice
field theory. Unfortunately, the energy conserving $\delta$ function in 
Eq.~(\ref{eq2.17}) prevents us from using these standard lattice techniques
for the Boltzmann equation. The reason is the following: When integrating over 
an arbitrary momentum mode in Eq.~(\ref{eq2.17}) one has to look for zeros of 
the argument of the energy conserving $\delta$ function with respect to this 
particular momentum 
mode. These zeros might well fall between two lattice sites. Hence, computing 
the collision integral requires the use of interpolation techniques in order 
to determine the particle number distribution for these in-between lattice 
sites. These interpolation techniques imply a continuity assumption for the 
particle number distribution which contradicts the strict lattice 
discretization as offered by lattice field theory. Apart from this principal
obstacle, there is also a practical reason which encourages us to use 
different discretization schemes for both types of equations: The collision 
integral in Eq.~(\ref{eq2.17}) is no convolution. Consequently, Fast Fourier
Transformation algorithms are not applicable, and its numerical computation
becomes rather expensive. In order to reduce the complexity of our Boltzmann 
numerics we exploited isotropy, which allowed us to simplify the Boltzmann
equation analytically and lead us to Eq.~(\ref{eq2.9}). In the discretized 
version of the Boltzmann equation (\ref{eq2.9}) the momenta are of the form
\[ p_n = \frac{\sqrt{12}}{a_s N_s} n \;. \]
We use the same value for $a_s$ as for the Kadanoff-Baym equations. This 
ensures that the largest available momentum is the same as for the 
Kadanoff-Baym equations. Of course, $N_s$ need not be the same as for the 
Kadanoff-Baym equations, which just means that we approach the physically
relevant infinite volume limit independently for both types of equations. 

In order to compute the collision integral we proceed as follows: For fixed 
$\left( k, p, q \right)$ we determine $r_0$ (the exact definition of $r_0$ is 
given in the appendix), which of course need not be one of the discretized 
momenta given above. The function $D \left( k, p, q, r_0 \right)$ can be 
evaluated for any value of $r_0$ (as one also can see in the appendix). To 
obtain the particle number density for an arbitrary $r_0$ we use a cubic 
spline interpolation~\cite{gsl}. Thus, for given $\left( k, p, q \right)$ the 
integrand is known to any given accuracy and for given $k$ we can simply sum 
over $p$ and $q$. In order to advance in time we use a Runge-Kutta-Cash-Karp 
routine with adaptive step-size control~\cite{gsl}. 

In order to set the scale for the simulations, again we use the renormalized 
vacuum mass $m_R$. Our simulations were done with $N_s = 500$, $a_s m_R = 
0.5$ and $\lambda = 18$.

\section{Comparing Boltzmann vs.\ Kadanoff-Baym}

\begin{figure*}[tb]
  \begin{minipage}{80mm}
    \includegraphics[width=80mm]{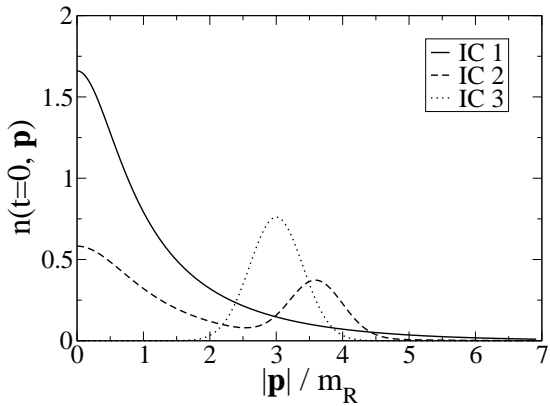}
  \end{minipage}
  \hfill
  \begin{minipage}{80mm}
    \caption{\label{fig2.6} {\bf Initial particle number densities}
    against absolute momenta. Shown are the three different initial 
    conditions (IC) discussed in the text, for which we numerically solved 
    the Boltzmann and the Kadanoff-Baym equations, respectively. All initial
    conditions correspond to the same (conserved) average energy density. 
    Above that, the initial conditions IC1 and IC2 also correspond to the 
    same initial average particle number density.}
  \end{minipage}
\end{figure*} 
We consider three different initial conditions which correspond to the same 
average energy density. Above that, the initial conditions IC1 and IC2 also
correspond to the same initial average particle number density. The 
corresponding initial particle number distributions are shown in 
Fig.~\ref{fig2.6}. These particle number distributions can immediately be fed 
into the numerics for the Boltzmann equation. In order to obtain the initial 
conditions for the Kadanoff-Baym equations, we follow 
Refs.~\cite{Berges:2001fi,Berges:2002wr}: The initial values for the spectral 
function are determined from the canonical commutation relations. On the other 
hand, for a given initial particle number density, the initial values for the 
statistical propagator and its derivatives are determined according to:
\begin{equation} \label{eq2.13}
  G_F \left( x^0, y^0, \bm{p} \right)_{x^0=y^0=0} = \left[ \frac{n \left( t, \bm{p} \right) + \frac{1}{2}}{\omega \left(t, \bm{p} \right)} \right]_{t=0} \;,
\end{equation}
\begin{equation} \label{eq2.14}
  \left[ \partial_{x^0} G_F \left( x^0, y^0, \bm{p} \right) \right]_{x^0=y^0=0} = 0 \;,
\end{equation}
\begin{equation} \label{eq2.15}
  \left[ \partial_{x^0} \partial_{y^0} G_F \left( x^0, y^0, \bm{p} \right) \right]_{x^0=y^0=0} = \left[ \omega \left(t, \bm{p} \right) \left( n \left( t, \bm{p} \right) + \frac{1}{2} \right) \right]_{t=0} \;,
\end{equation}
where the initial effective energy density is given by
\begin{figure*}[tb]
  \includegraphics[width=80mm]{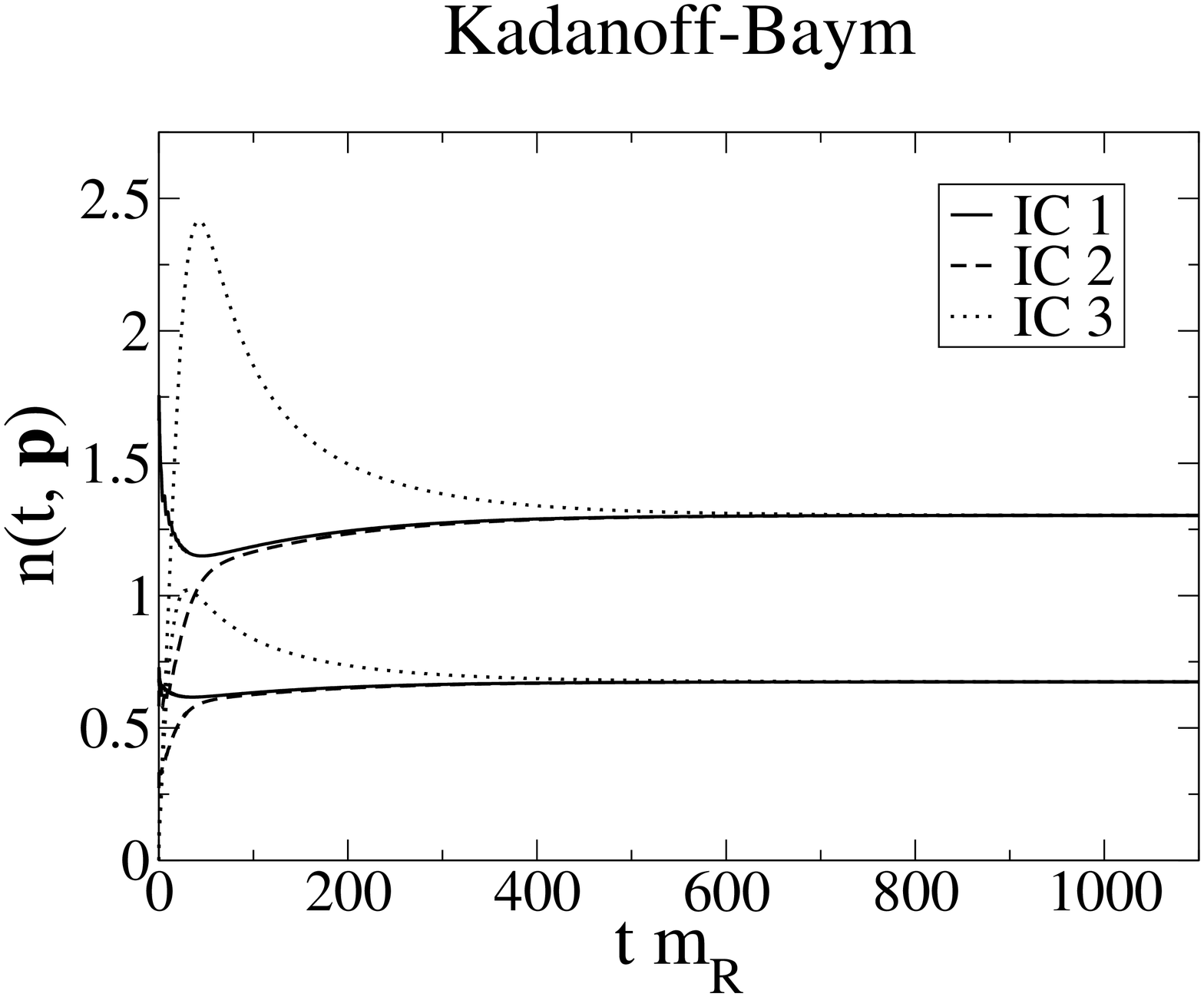}
  \hfill
  \includegraphics[width=80mm]{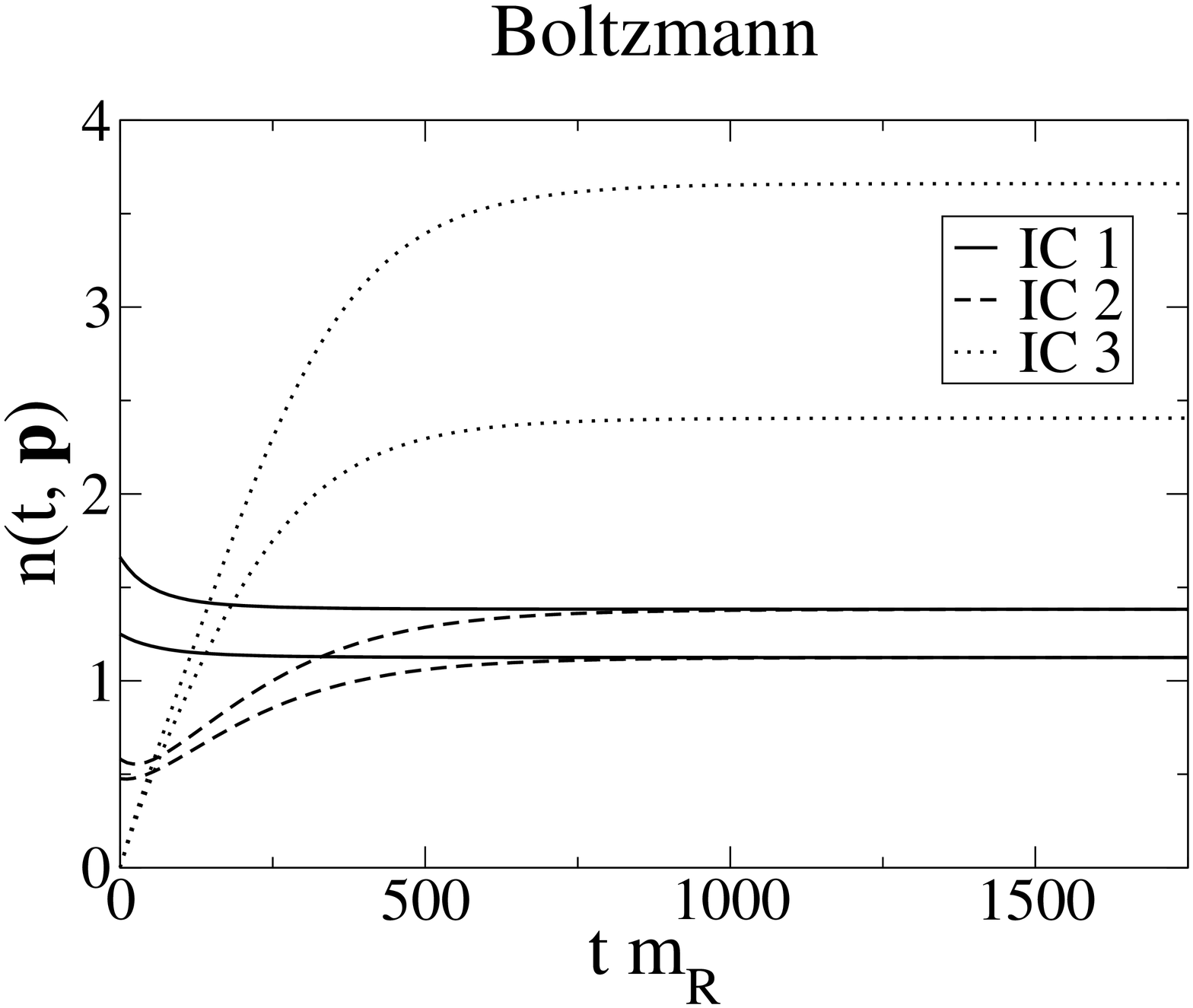}
  \caption{\label{fig2.8} These plots show the {\bf time evolution of the 
  particle number distributions} for two different momentum modes and all 
  initial conditions (cf.~Fig.~\ref{fig2.6}) as determined by the Boltzmann 
  and the Kadanoff-Baym equations, respectively. We see that the Kadanoff-Baym 
  equations respect full universality, whereas in the case of the Boltzmann 
  equation only a restricted universality is maintained, cf.\ 
  Fig.~\ref{fig2.9}.}
\end{figure*} 
\[ \omega \left( t=0, \bm{p} \right) = \sqrt{m_R^2 + \bm{p}^2} \;. \]
Figs.~\ref{fig2.8} and \ref{fig2.9} show the evolution of the particle number 
distributions for two momentum modes and the corresponding equilibrium 
particle number distributions, respectively, for all initial conditions. In 
the left plots we can see, that the Kadanoff-Baym equations lead to a 
universal equilibrium particle number density. The left plot in 
Fig.~\ref{fig2.8} shows that the particle number distributions may evolve 
quite differently for early times\footnote{As we will see, the steep 
over-shooting of the particle number distribution leads to a quick kinetic 
equilibration, whereas the rather long tail accounts for chemical 
equilibration.}. However, respecting universality, for any given momentum mode
all distributions approach the same late-time value. This plot is supplemented 
by the left plot in Fig.~\ref{fig2.9}. There, one can see that the various 
particle number densities, after equilibrium has effectively been reached, 
indeed completely agree. Hence, this plot proves that we could have shown 
plots similar to the left one in Fig.~\ref{fig2.8} for all momentum modes. 
In particular the predicted temperature, given by the inverse slope of the 
line, is the same for all initial conditions. In contrast to this, the right 
plots reveal that the Boltzmann equation respects only a restricted 
universality. In general, e.g.~for the initial conditions IC1 and IC3, for 
any given momentum mode the particle number densities will not approach the 
same late-time value. For both momentum modes shown in Fig.~\ref{fig2.8} a 
considerable discrepancy is revealed. However, for the special case of the 
initial conditions IC1 and IC2, which, as mentioned above, correspond to the 
same initial average particle number density, the late-time results do 
agree\footnote{In Fig.~\ref{fig2.9} one can see that in the case of the 
Boltzmann equation there is only one momentum mode for which the late-time 
values of all particle number densities agree, namely the intersection point 
of the lines. However, we could easily have chosen a fourth initial condition 
for which the late-time result would intersect the lines in Fig.~\ref{fig2.9} 
in different points. Then there would not be a single momentum mode for which 
the late-time values of all particle number densities agreed.}.

The reason for the observed restriction of universality can be extracted from 
Fig.~\ref{fig2.10}. There we show the time evolution of the total particle 
number per volume
\[ N_{tot} \left( t \right) = \int \frac{\ddd{p}}{\left( 2 \pi \right)^3} \; n \left( t, \bm{p} \right) \;. \]
In general the Kadanoff-Baym equations conserve the average energy 
density\footnote{Concerning our simulations, of course, this only holds up to 
numerical errors. We have checked that our simulations conserve the average 
energy density up to a numerical uncertainty of $0.2 \%$ for the Kadanoff-Baym
equations and the Boltzmann equation, respectively.} and global 
charges~\cite{baymKadanoff1961a,Baym:1962sx,Ivanov:1998nv}. 
\begin{figure*}[tb]
  \includegraphics[width=80mm]{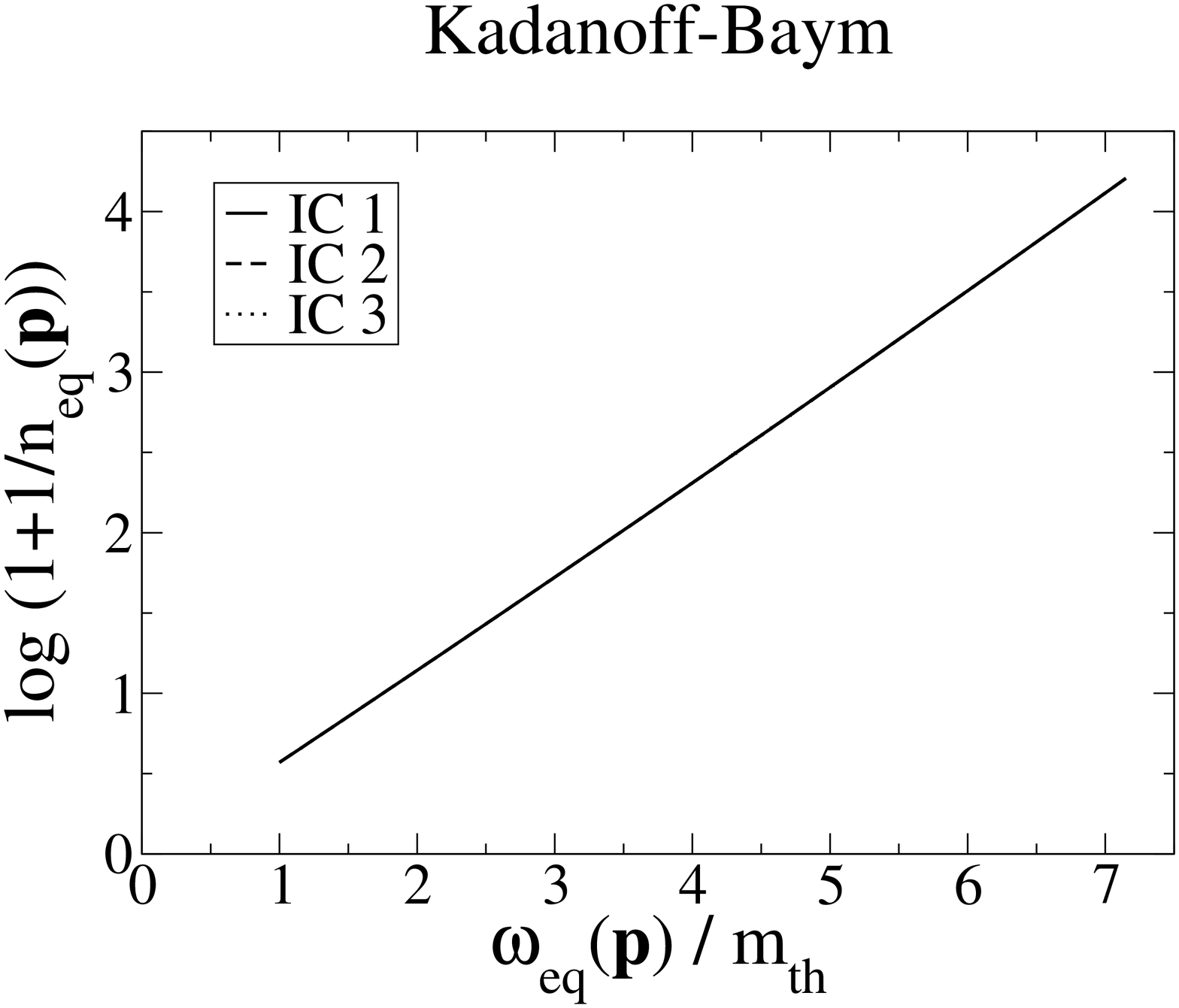}
  \hfill
  \includegraphics[width=80mm]{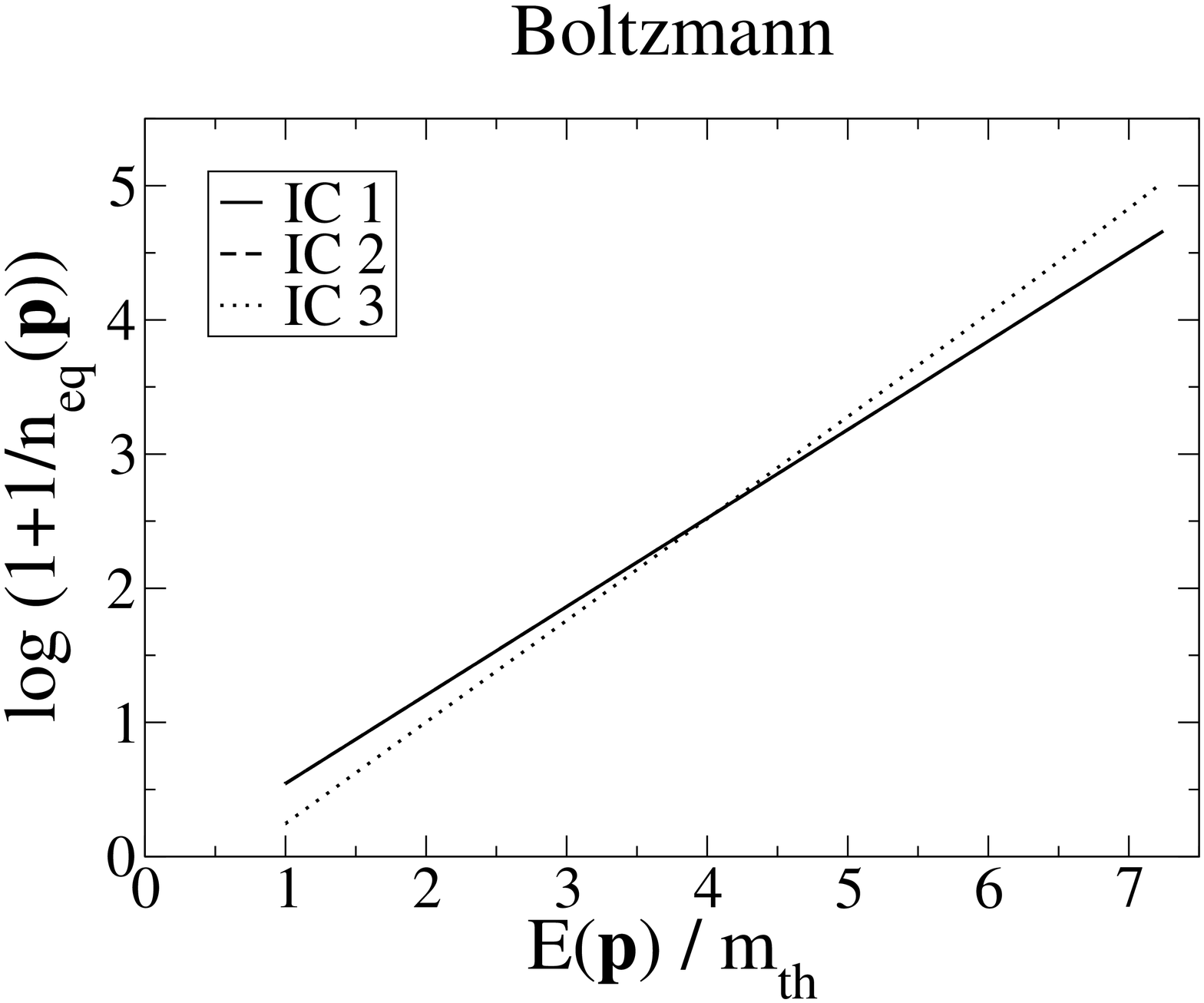}
  \caption{\label{fig2.9} Here, we plotted the {\bf equilibrium particle 
  number distributions} obtained for times when thermal equilibrium has 
  effectively been reached, against the corresponding thermal energy densities.
  The thermal mass is given by the zero mode of the effective kinetic 
  equilibrium energy density as determined by the Kadanoff-Baym equations: 
  $m_{th} = \omega_{eq} \left( \bm{p}=0 \right)$. For a given initial 
  condition, the temperature is given by the inverse slope of the line and the 
  chemical potential is obtained from the y-axis intercept divided by 
  $- \beta$. Supplementing Fig.~\ref{fig2.8} we observe full (restricted) 
  universality in 
  the case of the Kadanoff-Baym (Boltzmann) equations. In particlular, the 
  Kadanoff-Baym equations lead to a universal temperature $T = 1.68 \; 
  m_{th}$ and a universally vanishing chemical potential. In contrast to this, 
  the Boltzmann equation gives $T = 1.52 \; m_{th}$ and $\mu = 0.18 \; 
  m_{th}$ for the initial conditions IC1 and IC2, and $T = 1.32 m_{th}$ and 
  $\mu = 0.68 \; m_{th}$ for IC3.}
\end{figure*} 
However, as there is no conserved charge in our theory, the total particle 
number need not be conserved. Indeed, the Kadanoff-Baym equations include 
off-shell particle creation and annihilation~\cite{Aarts:2001qa}. 
Consequently, the total particle number may change, and in fact approaches 
a universal equilibrium value. In contrast to this, due to the quasi-particle 
(or on-shell) approximation particle number changing processes are 
kinematically forbidden in the Boltzmann equation. The Boltzmann equation 
only includes two-particle scattering, which leaves the total particle number 
constant. Of course, 
this additional constant of motion severely restricts the evolution of the 
particle number density. Therefore the Boltzmann equation cannot lead to a 
universal quantum thermal equilibrium. Only initial conditions for which the 
average energy density and the total particle number agree from the very 
beginning, lead to the same equilibrium results.

In a system allowing for creation and annihilation of particles, the chemical 
potential of particles, whose total number is not restricted by any conserved 
quantity, must vanish in thermodynamical equilibrium. The chemical potential
predicted by the Kadanoff-Baym and Boltzmann equations, respectively, is given
by the y-axis intercept, extracted from Fig.~\ref{fig2.9}, divided by 
$- \beta$. Using a ruler the reader might convince himself that the 
Kadanoff-Baym equations indeed lead to a universally vanishing chemical 
potential. 
\begin{figure*}[tb]
  \includegraphics[width=80mm]{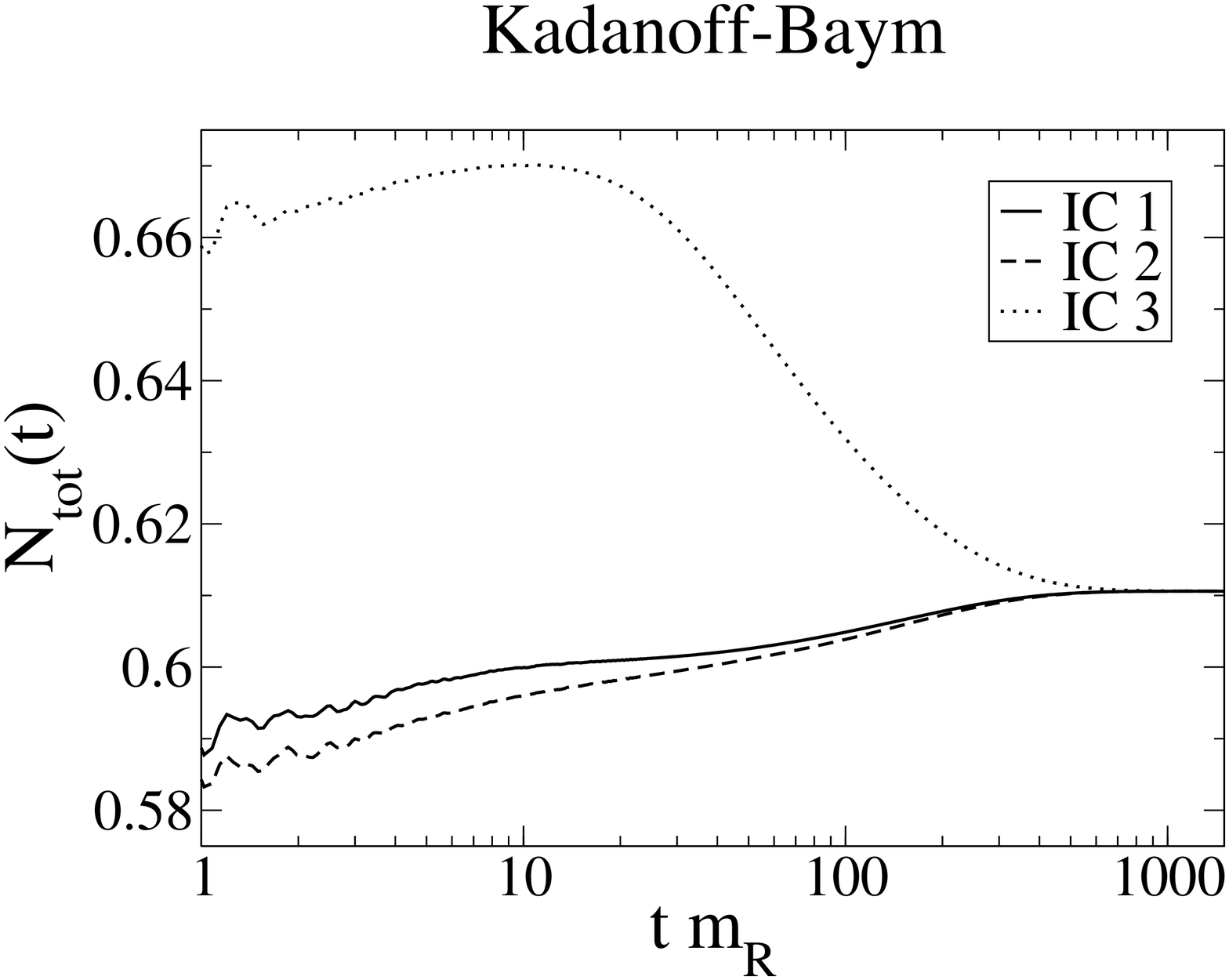}
  \hfill
  \includegraphics[width=80mm]{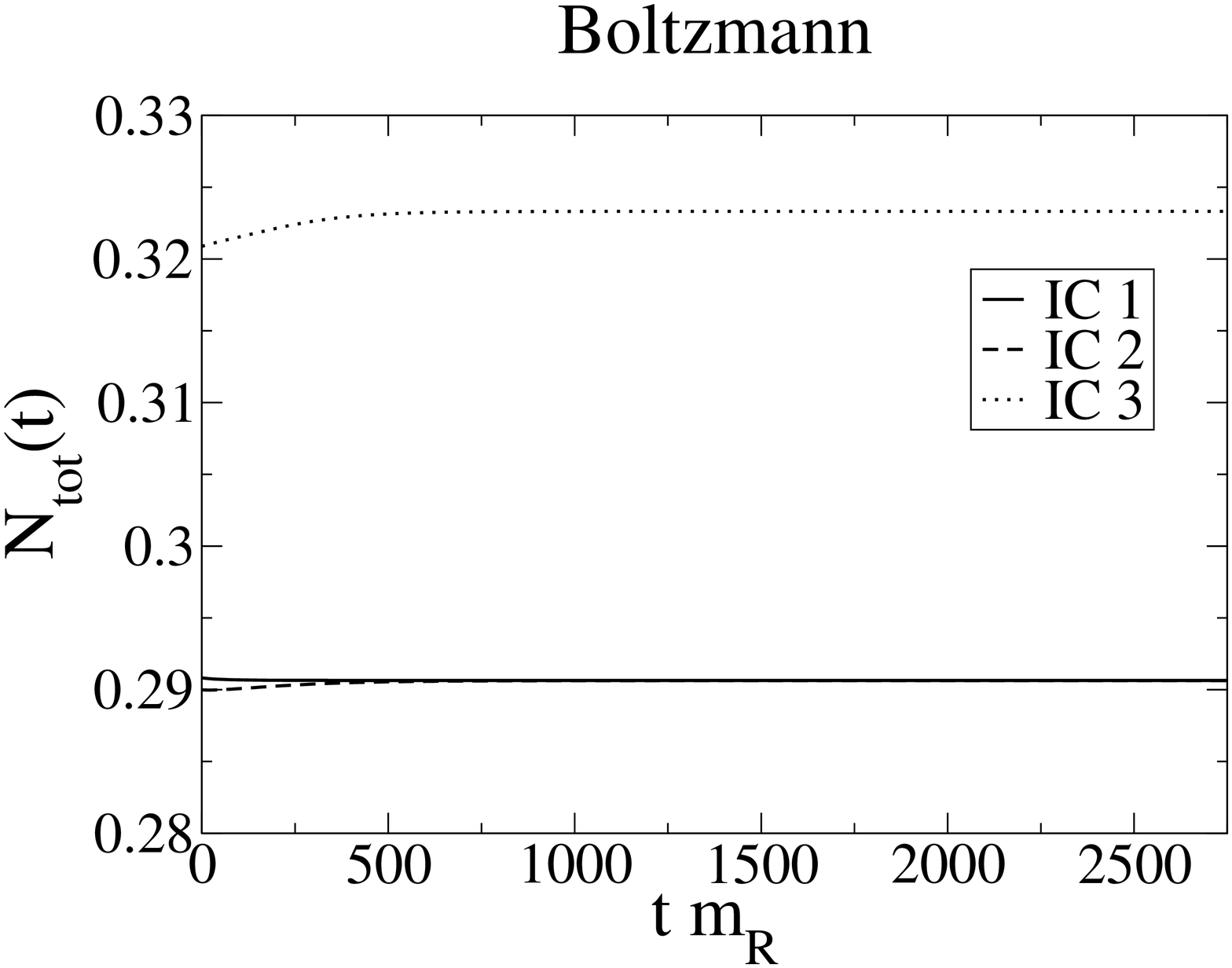}
  \caption{\label{fig2.10} {\bf Time evolution of the total particle number}.
  As expected from Ref.~\cite{Aarts:2001qa}, the Kadanoff-Baym equations 
  include off-shell particle creation and annihilation. As a result the total 
  particle number may change with time. In contrast to this the total particle
  number is strictly conserved in the case of the Boltzmann 
  equation. Concerning our simulations, of course, this only holds up to 
  numerical errors ($< 0.8 \%$). The quantitative disagreement of the total
  particle numbers in both plots can be attributed to the substantial 
  discrepancies in the discretization schemes underlying our Boltzmann and 
  Kadanoff-Baym numerics and are of no relevance for the purposes of the 
  present work.}
\end{figure*} 
In contrast to this, even without a ruler one can see that the 
Boltzmann equation, in general, will lead to a non-vanishing chemical 
potential\footnote{For the initial conditions considered in this work, the 
Boltzmann equation predicted even a positive chemical potential. However, 
already on very general grounds, one can deduce that the chemical potential 
of bosons has to be negative!}.

\begin{figure*}[tb]
  \includegraphics[width=80mm]{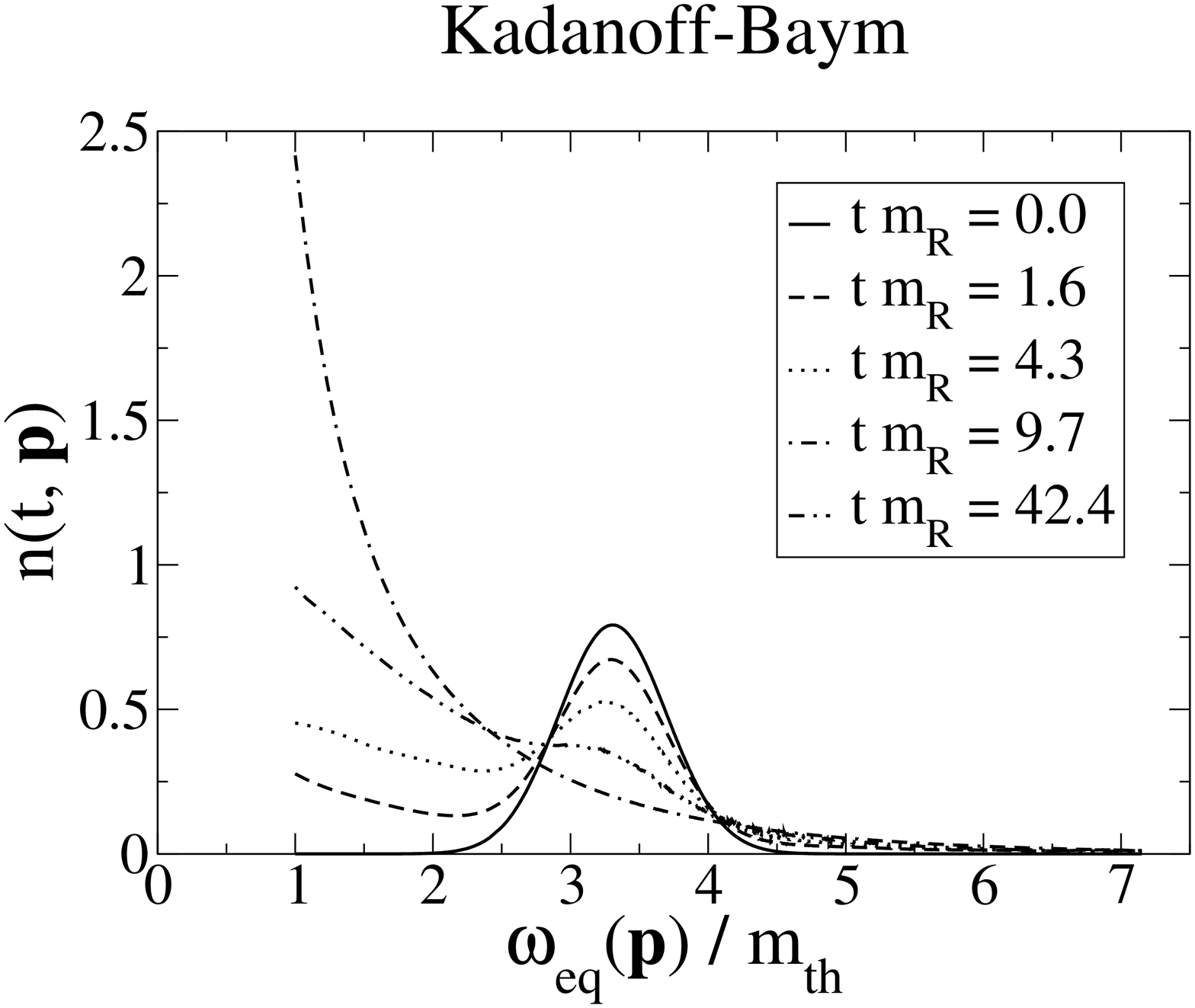}
  \hfill
  \includegraphics[width=80mm]{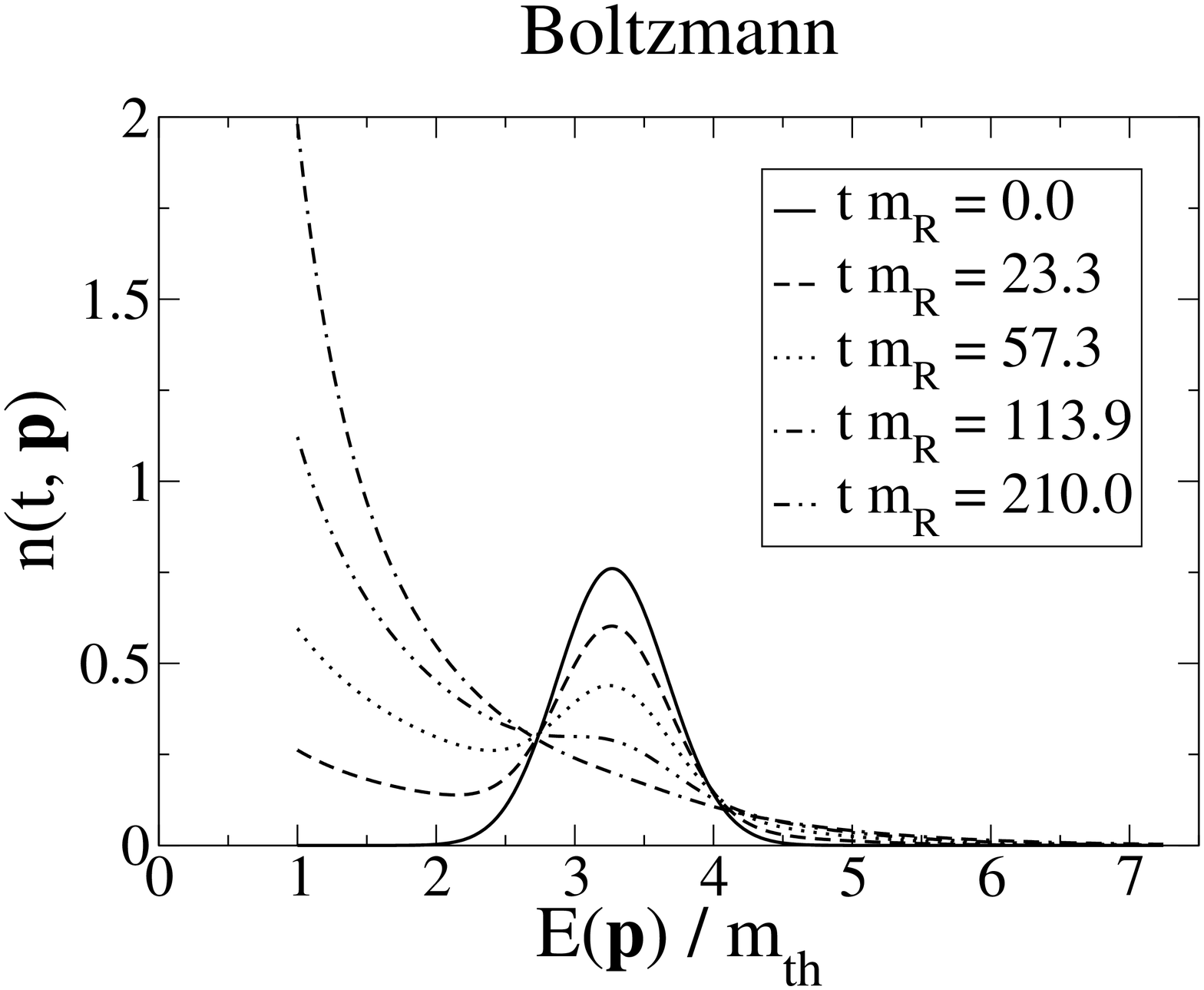} \\
  \includegraphics[width=80mm]{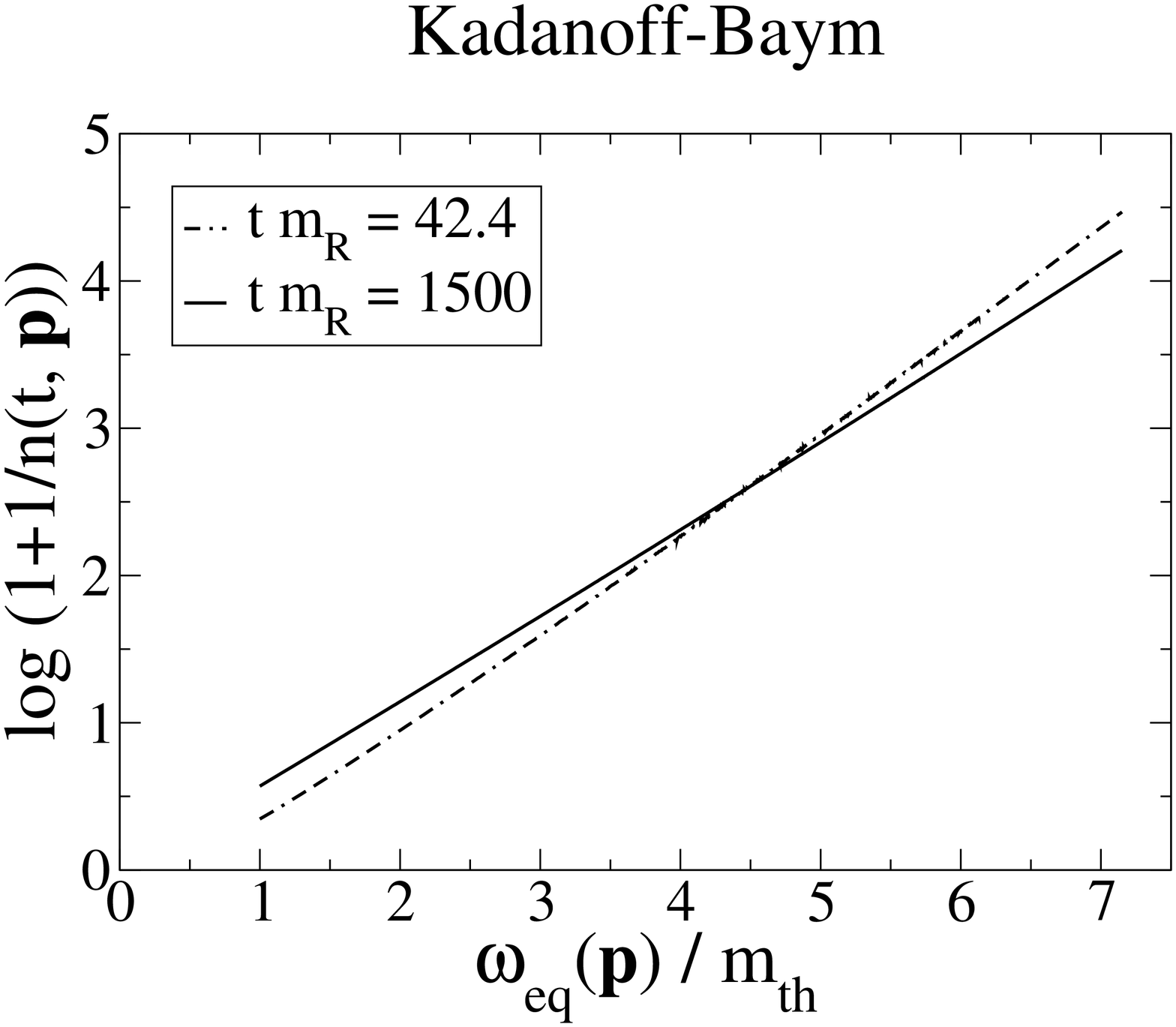}
  \hfill
  \includegraphics[width=80mm]{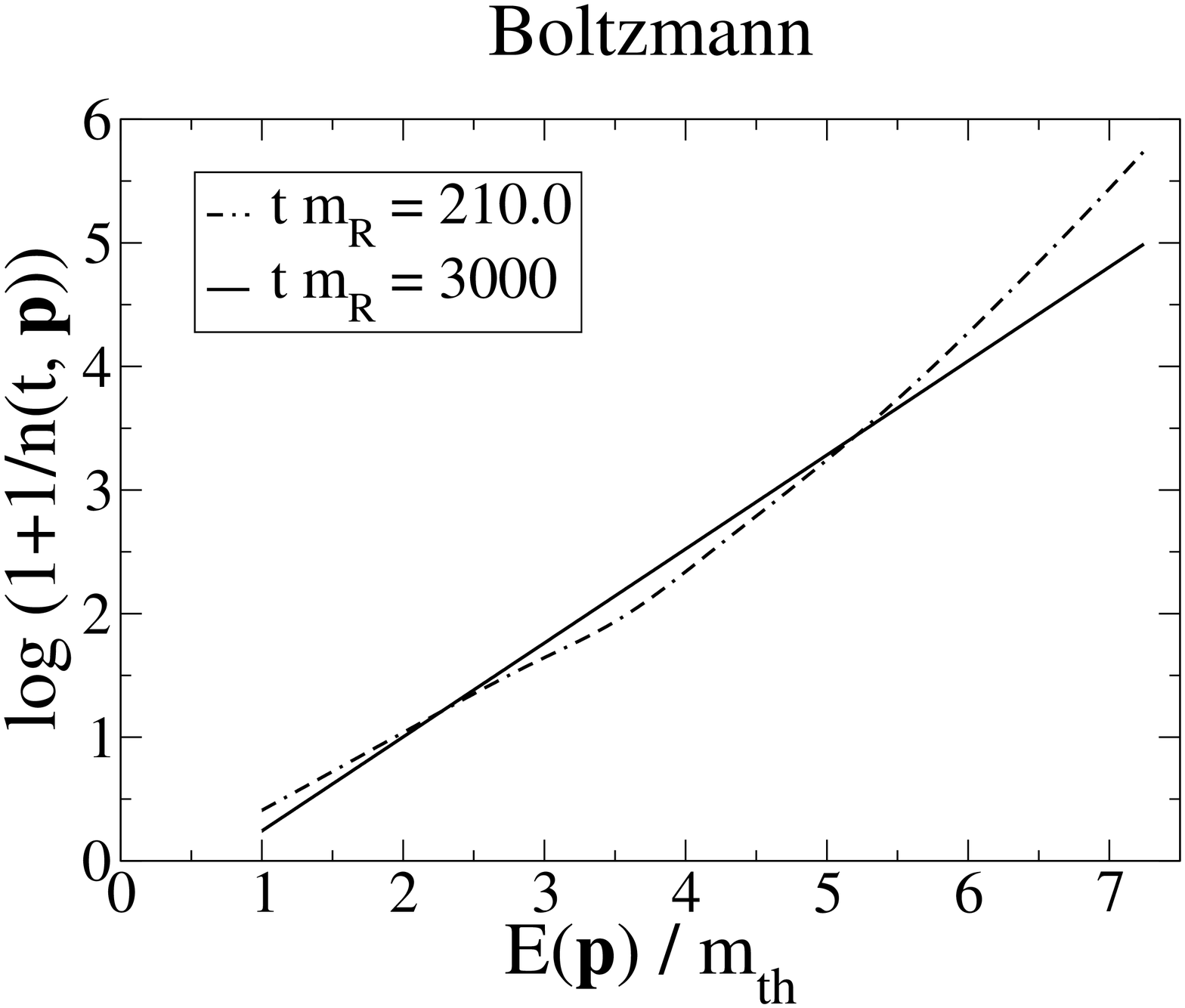} \\
  \caption{\label{fig2.12} {\bf (Missing) separation of time scales.}
  The particle number distribution is shown against the equilibrium energy 
  density at various times for initial condition IC3.}
\end{figure*}

In this context, Fig.~\ref{fig2.12} exhibits further interesting results. In 
the upper left plot one can see that the Kadanoff-Baym equations rapidly wash
out our tsunami-type initial condition IC3. In both plots on the left hand side
the double-dashed-dotted lines correspond to the particle number distribution 
at the same time $t m_R = 42.4$. Thus, in the lower left plot one obtains an 
approximate straight line
already after a relatively short period of time, indicating a swift approach 
to kinetic equilibrium. Subsequently, this straight line is tilted until it 
intersects the origin of our coordinate system (full line), corresponding to a
vanishing chemical potential . However, this approach to full thermodynamical
(including chemical) equilibrium takes a considerably longer 
time~\cite{Berges:2004ce}. In this way, the left plots reveal two distinct 
time scales: a rather fast kinetic equilibration, and a very slow 
thermodynamical (including chemical) equilibration. These two time scales can 
also be identified in the left plot of Fig.~\ref{fig2.8} and in 
Fig.~\ref{fig2.4}. The over-shooting of the particle number density for early 
times leads to the kinetic equilibration. In fact, the double-dashed-dotted 
lines correspond to the time, when the particle number distribution 
(equal-time propagator) reaches its maximum value in Fig.~\ref{fig2.8}
(Fig.~\ref{fig2.4}). Interestingly, although the initial conditions IC1 and 
IC2 do not show this excessive over-shooting, the corresponding particle 
number distributions (equal-time propagators) approach each other on the same 
time scale, from which on they show an almost identical evolution. The 
following rather long tail, again indicates that full thermalization takes 
place on much larger time scales.
The right plot in Fig.~\ref{fig2.8} shows that the steep initial evolution,
which is characteristic for the Kadanoff-Baym equations, is absent in the case
of the Boltzmann equation\footnote{One might be tempted to conclude that the 
evolution of the particle number distribution is strictly monotonous in the 
Boltzmann case. However, the small dip for the particle number distribution 
IC2 in Fig.~\ref{fig2.8} shows that this is not necessarily the case.} and 
that the Boltzmann equation leads only to a gently inclined evolution of the 
particle number distribution.
Accordingly, the plots on the right hand side of Fig.~\ref{fig2.12} show that 
it takes a considerably longer time for the Boltzmann equation to reach 
kinetic equilibrium. As already mentioned above, in contrast to the 
Kadanoff-Baym equations, the Boltzmann equation cannot describe the process 
of chemical equilibration. Consequently, the separation of time scales 
furnished by the Kadanoff-Baym equations is absent in the Boltzmann case.

\section{Conclusions}

Starting from the 2PI effective action for a scalar $\Phi^4$ quantum field
theory we briefly reviewed the derivation of the Kadanoff-Baym equations and
the approximations which are necessary to eventually arrive at a Boltzmann 
equation. We solved both, the Kadanoff-Baym equations and the Boltzmann 
equation, numerically for spatially homogeneous and isotropic systems in 3+1
dimensions without any further approximations.

We have shown that the Kadanoff-Baym equations respect universality: For
systems with equal average energy density the late time behavior coincides
independent of the details of the initial conditions. In particular, 
independent of the initial conditions the particle number densities, 
temperatures and thermal masses predicted for times when equilibrium has 
effectively been reached coincide. The chemical potentials also coincide and 
vanish. Furthermore, we observed that thermalization takes place on two 
different time-scales: a rather fast kinetic equilibration, and a very slow 
thermodynamical (including chemical) equilibration.

In general Kadanoff-Baym and Boltzmann equations conserve total energy as well
as global charges. In the special case of a real scalar $\Phi^4$ quantum field 
theory the quasi-particle approximation implies that the Boltzmann equation 
additionally
conserves the total particle number. This additional constant of motion 
severely restricts the evolution of the system. As a result the Boltzmann 
equation cannot lead to a universal quantum thermal equilibrium. The Boltzmann
equation respects only a restricted universality: Only initial conditions for 
which the average energy density and the total particle number agree from the 
very beginning, lead to the same equilibrium results. In particular, the 
Boltzmann equation cannot describe the phenomenon of chemical equilibration 
and, in general, will lead to a non-vanishing chemical potential. Due to the 
lack of chemical equilibration, the separation of time scales, which we 
observed for the Kadanoff-Baym equations, is absent in the case of the 
Boltzmann equation.

Some of the approximations that lead from the Kadanoff-Baym equations to the
Boltzmann equation (namely, the gradient expansion, neglecting the Poisson
brackets and the Kadanoff-Baym ansatz) are clearly motivated by equilibrium
considerations. Taking the observed restriction of universality into account,
we conclude that one can safely apply the Boltzmann equation only to systems 
which are sufficiently close to equilibrium. Accordingly, for a system far 
from equilibrium the results given by the Boltzmann equation should be 
treated with care.

In the future it will be important to perform a similar comparison between
Boltzmann and Kadanoff-Baym equations for Yukawa-type quantum field theories 
including fermions and gauge fields. Also a treatment of Kadanoff-Baym 
equations on an expanding space-time should reveal interesting results. This 
would finally enable one to attack the problem of leptogenesis. Independent of
the comparison between Boltzmann and Kadanoff-Baym equations we are looking 
forward to learn to which extend a full non-perturbative renormalization 
procedure~\cite{vanHees:2001ik,vanHees:2001pf,vanHees:2002bv,Blaizot:2003an,
Berges:2004hn,Berges:2005hc} affects the results quantitatively.

\begin{acknowledgments}
  TripleM would like to thank J\"urgen Berges for many fruitful discussions 
  and collaboration on related work. Furthermore we would like to thank, 
  Mathias Garny, Patrick Huber and Andreas Hohenegger for discussions and 
  valuable hints. This work was supported by the  ``Sonderforschungsbereich 
  375 f\"ur Astroteilchenphysik der Deutschen Forschungsgemeinschaft''.
\end{acknowledgments}

\appendix*

\section*{Simplifying the Boltzmann Equation}

The simplification of the Boltzmann equation~\cite{Dolgov:1997mb}
relies on the Fourier representation of the momentum conservation delta 
function:
\[ \delta^3 \left( \bm{m} \right) = \int \frac{\ddd{\xi}}{\left( 2 \pi \right)^3} \; \exp \left( - i \bm{m} \bm{\xi} \right) \;. \]
Using spherical coordinates, we find
\[ \bm{m} \bm{\xi} = m \xi \Big( \sin \vartheta_m \; \sin \vartheta_{\xi} \; \cos \left( \varphi_m - \varphi_{\xi} \right) + \cos \vartheta_m \; \cos \vartheta_{\xi} \Big) \;. \]
Now, we consider just the integration over the solid angle. As we integrate 
over the complete solid angle $\Omega_{\xi}$, it does not matter in which 
direction $\bm{m}$ is pointing. The result will always be the same:
\[ \int d\Omega_{\xi} \exp \left( - i \bm{m} \bm{\xi} \right) = \int d\Omega_{\xi} \exp \left( - i \bm{m}_0 \bm{\xi} \right) \;, \]
where we can choose $\bm{m}_0 = \left( 0, 0, m \right)$, such that $\varphi_m 
= \vartheta_m = 0$. Now, we can evaluate the integral quite easily:
\begin{equation} \label{eq2.19}
  \int d\Omega_{\xi} \exp \left( - i \bm{m} \bm{\xi} \right) = \int d\Omega_{\xi} \exp \left( - i m \xi \cos \vartheta_{\xi} \right) = \frac{4 \pi}{m \xi} \sin \left( m \xi \right) \;.
\end{equation}
After we have rewritten Eq.~(\ref{eq2.17}) using spherical coordinates and
inserted the Fourier representation for the momentum conservation delta 
function, we can use Eq.~(\ref{eq2.19}) to perform the integrations over the 
solid angles. Here it is crucial first to evaluate the integrals over 
$\Omega_p$, $\Omega_q$ and $\Omega_r$, and to do the integral over 
$\Omega_{\xi}$ at last. We find:
\begin{eqnarray*}
        \lefteqn{\partial_t n \left( t, k \right) = \frac{\lambda^2}{96 \pi^4} \intl_0^{\infty} dp \intl_0^{\infty} dq \intl_0^{\infty} dr \intl_0^{\infty} d\xi \Bigg[ p q r \frac{\delta \left( E_k + E_p - E_q - E_r \right)}{E_k E_p E_q}} \\
  &   & {} \times \frac{1}{k \xi^2} \sin \left( k \xi \right) \; \sin \left( p \xi \right) \; \sin \left( q \xi \right) \; \sin \left( r \xi \right) \\
  &   & {} \times \Big( \left( 1 + n_k \right) \left( 1 + n_p \right) n_q n_r - n_k n_p \left( 1 + n_q \right) \left( 1 + n_r \right) \Big) \Bigg] \;,
\end{eqnarray*}
There are only two more steps to make in order to arrive at Eq.~(\ref{eq2.9}).
First, we define the auxiliary function $D \left( k, p, q, r \right)$:
\[ D \left( k, p, q, r \right) = \intl_0^{\infty} d\xi \frac{1}{k \xi^2} \sin \left( k \xi \right) \; \sin \left( p \xi \right) \; \sin \left( q \xi \right) \; \sin \left( r \xi \right) \;. \]
This can easily be evaluated using a computer algebra program. For $k > 0$ 
this is 
\begin{eqnarray*}
        D \left( k, p, q, r \right)
  & = & \frac{\pi}{16 k} \Big( \left| k - p - q - r \right| - \left| k + p - q - r \right| \\
  &   & {} - \left| k - p + q - r \right| + \left| k + p + q - r \right| \\
  &   & {} - \left| k - p - q + r \right| + \left| k + p - q + r \right| \\
  &   & {} + \left| k - p + q + r \right| - \left| k + p + q + r \right| \Big) \;,
\end{eqnarray*}
and for $k = 0$ we obtain
\begin{eqnarray*}
        D \left( 0, p, q, r \right)
   &=&  \frac{\pi}{8} \Big( \sign \left( p + q - r \right) - \sign \left( p - q - r \right) \\
   & & {} + \sign \left( p - q + r \right) - \sign \left( p + q + r \right) \Big).
\end{eqnarray*}
Second, we use the energy conservation $\delta$ function to evaluate the 
integral over $r$, using the well-known formula 
\[ \delta \left( f \left( r \right) \right) = \sum_{\left\{r_0 | f \left( r_0 \right) = 0 \right\}} \frac{\delta \left( r - r_0 \right)}{\left| \left( \frac{df}{dr} \right)_{r=r_0} \right|} \;. \]
$r_0$ is determined by the condition that the argument of the energy 
conservation $\delta$ function is zero:
\begin{equation} \label{eq2.20}
  E_k + E_p - E_q - E_{r_0} = 0 \;.
\end{equation}
If this condition can be satisfied, $r_0$ is given by
\[ r_0 = r_0 \left( t, k, p, q \right) = \sqrt{\left( E_k + E_p - E_q \right)^2 - M^2 \left( t \right)} \;. \]
If $k$, $p$ and $q$ are such that condition (\ref{eq2.20}) cannot be 
satisfied, the above square root yields a purely imaginary result and 
$r_0^2 < 0$. Due to the $\Theta$ function the corresponding term does not
contribute to the collision integral. After these final steps we end up 
exactly with Eq.~(\ref{eq2.9}).


\input{scalars.bbl}
\end{document}

%% file: scalarsInput.tex
\newcommand{\TUM}{Physik-Department T30d, Technische Universit\"at M\"unchen\\
                  James-Franck-Stra\3e, 85748 Garching, Germany}

\newcommand{\MPPMU}{Max-Planck-Institut f\"ur Physik 
                    (Werner-Heisenberg-Institut)\\
                    F\"ohringer Ring 6, 80805 M\"unchen, Germany}

\renewcommand{\c}{\mathcal{C}}

\newcommand{\ddd}[1]{d^3 #1}

\newcommand{\dddd}[1]{d^4 #1}

\newcommand{\intl}{\int\limits}

\renewcommand{\Re}{\mbox{Re}}

\newcommand{\sign}{\mbox{sign}}

\newcommand{\tr}{\mbox{tr}}


%% file: scalars.bbl
\begin{thebibliography}{10}
\expandafter\ifx\csname url\endcsname\relax
  \def\url#1{\texttt{#1}}\fi
\expandafter\ifx\csname urlprefix\endcsname\relax\def\urlprefix{URL }\fi
\providecommand{\eprint}[2][]{\url{#2}}

\bibitem{Fukugita:1986hr}
M.~Fukugita and T.~Yanagida, \emph{Baryogenesis without Grand Unification},
  Phys. Lett.
\textbf{B174} (1986) 45.

\bibitem{Buchmuller:2004nz}
W.~Buchm{\"u}ller, P.~Di~Bari, and M.~Pl{\"u}macher, \emph{Leptogenesis for
  pedestrians}, Ann. Phys. \textbf{315} (2005) 305,
\eprint{hep-ph/0401240}.

\bibitem{Buchmuller:2000nd}
Wilfried Buchm{\"u}ller and Stefan Fredenhagen, \emph{Quantum mechanics of
  baryogenesis}, Phys. Lett. \textbf{B483} (2000) 217,
\eprint{hep-ph/0004145}.

\bibitem{Sakharov:1967dj}
A.~D. Sakharov, \emph{Violation of CP Invariance, C Asymmetry, and Baryon
  Asymmetry of the Universe}, JETP Lett.
\textbf{5} (1967) 24.

\bibitem{Arsene:2004fa}
I.~Arsene et~al. (BRAHMS), \emph{Quark gluon plasma and color glass condensate
  at RHIC? The perspective from the BRAHMS experiment}, Nucl. Phys.
  \textbf{A757} (2005) 1,
\eprint{nucl-ex/0410020}.

\bibitem{Back:2004je}
B.~B. Back et~al. (PHOBOS), \emph{The PHOBOS perspective on discoveries at
  RHIC}, Nucl. Phys. \textbf{A757} (2005) 28,
\eprint{nucl-ex/0410022}.

\bibitem{Adams:2005dq}
J.~Adams et~al. (STAR), \emph{Experimental and theoretical challenges in the
  search for the quark gluon plasma: The STAR collaboration's critical
  assessment of the evidence from RHIC collisions}, Nucl. Phys. \textbf{A757}
  (2005) 102,
\eprint{nucl-ex/0501009}.

\bibitem{Adcox:2004mh}
K.~Adcox et~al. (PHENIX), \emph{Formation of dense partonic matter in
  relativistic nucleus nucleus collisions at RHIC: Experimental evaluation by
  the PHENIX collaboration}, Nucl. Phys. \textbf{A757} (2005) 184,
\eprint{nucl-ex/0410003}.

\bibitem{Berges:2004ce}
J.~Berges, S.~Bors{\'a}nyi, and C.~Wetterich, \emph{Prethermalization}, Phys.
  Rev. Lett. \textbf{93} (2004) 142002,
\eprint{hep-ph/0403234}.

\bibitem{Mueller:2002gd}
A.~H. Mueller and D.~T. Son, \emph{On the equivalence between the Boltzmann
  equation and classical field theory at large occupation numbers}, Phys. Lett.
  \textbf{B582} (2004) 279,
\eprint{hep-ph/0212198}.

\bibitem{Jeon:2004dh}
Sangyong Jeon, \emph{The Boltzmann equation in classical and quantum field
  theory}, Phys. Rev. \textbf{C72} (2005) 014907,
\eprint{hep-ph/0412121}.

\bibitem{baymKadanoff1962a}
Gordon Baym and Leo~P. Kadanoff, \emph{Quantum Statistical Mechanics}
  (Benjamin, New York, 1962)

\bibitem{Danielewicz:1982kk}
P.~Danielewicz, \emph{Quantum Theory of Nonequilibrium Processes I}, Annals
  Phys.
\textbf{152} (1984) 239.

\bibitem{Ivanov:1999tj}
Yu.~B. Ivanov, J.~Knoll, and D.~N. Voskresensky, \emph{Resonance Transport and
  Kinetic Entropy}, Nucl. Phys. \textbf{A672} (2000) 313,
\eprint{nucl-th/9905028}.

\bibitem{Knoll:2001jx}
J.~Knoll, Yu.~B. Ivanov, and D.~N. Voskresensky, \emph{Exact Conservation Laws
  of the Gradient Expanded Kadanoff- Baym Equations}, Annals Phys. \textbf{293}
  (2001) 126,
\eprint{nucl-th/0102044}.

\bibitem{Blaizot:2001nr}
Jean-Paul Blaizot and Edmond Iancu, \emph{The quark-gluon plasma: Collective
  dynamics and hard thermal loops}, Phys. Rept. \textbf{359} (2002) 355,
\eprint{hep-ph/0101103}.

\bibitem{Berges:2001fi}
J{\"u}rgen Berges, \emph{Controlled nonperturbative dynamics of quantum fields
  out of equilibrium}, Nucl. Phys. \textbf{A699} (2002) 847,
\eprint{hep-ph/0105311}.

\bibitem{Aarts:2001qa}
Gert Aarts and J{\"u}rgen Berges, \emph{Nonequilibrium time evolution of the
  spectral function in quantum field theory}, Phys. Rev. \textbf{D64} (2001)
  105010,
\eprint{hep-ph/0103049}.

\bibitem{Kolb:1979qa}
Edward~W. Kolb and Stephen Wolfram, \emph{Baryon Number Generation in the Early
  Universe}, Nucl. Phys.
\textbf{B172} (1980) 224.

\bibitem{kolbTurner1990a}
Edward~W. Kolb and Michael~S. Turner, \emph{The Early Universe}
  (Addison-Wesley, 1990)

\bibitem{kohler1995a}
H.~S. K{\"o}hler, \emph{Memory and correlation effects in nuclear collisions},
  Phys. Rev. \textbf{C51} (1995) 3232

\bibitem{Morawetz:1998em}
K.~Morawetz and H.~S. K{\"o}hler, \emph{Formation of correlations and
  energy-conservation at short time scales}, Eur. Phys. J. \textbf{A4} (1999)
  291,
\eprint{nucl-th/9802082}.

\bibitem{Juchem:2003bi}
S.~Juchem, W.~Cassing, and C.~Greiner, \emph{Quantum dynamics and
  thermalization for out-of-equilibrium $\phi^4$-theory}, Phys. Rev.
  \textbf{D69} (2004) 025006,
\eprint{hep-ph/0307353}.

\bibitem{baymKadanoff1961a}
Gordon Baym and Leo~P. Kadanoff, \emph{Conservation Laws and Correlation
  Functions}, Phys. Rev. \textbf{124} (1961) 287

\bibitem{Baym:1962sx}
Gordon Baym, \emph{Selfconsistent approximation in many body systems}, Phys.
  Rev.
\textbf{127} (1962) 1391.

\bibitem{Ivanov:1998nv}
Yu.~B. Ivanov, J.~Knoll, and D.~N. Voskresensky, \emph{Self-consistent
  approximations to non-equilibrium many-body theory}, Nucl. Phys.
  \textbf{A657} (1999) 413,
\eprint{hep-ph/9807351}.

\bibitem{Jackiw:1974cv}
R.~Jackiw, \emph{Functional evaluation of the effective potential}, Phys. Rev.
\textbf{D9} (1974) 1686.

\bibitem{Cornwall:1974vz}
John~M. Cornwall, R.~Jackiw, and E.~Tomboulis, \emph{Effective Action for
  Composite Operators}, Phys. Rev.
\textbf{D10} (1974) 2428.

\bibitem{Calzetta:1986cq}
E.~Calzetta and B.~L. Hu, \emph{Nonequilibrium quantum fields: Closed-time-path
  effective action, Wigner function and Boltzmann equation}, Phys. Rev.
\textbf{D37} (1988) 2878.

\bibitem{Berges:2000ur}
J{\"u}rgen Berges and J{\"u}rgen Cox, \emph{Thermalization of quantum fields
  from time-reversal invariant evolution equations}, Phys. Lett. \textbf{B517}
  (2001) 369,
\eprint{hep-ph/0006160}.

\bibitem{Berges:2002wr}
J{\"u}rgen Berges, Szabolcs Bors{\'a}nyi, and Julien Serreau,
  \emph{Thermalization of fermionic quantum fields}, Nucl. Phys. \textbf{B660}
  (2003) 51,
\eprint{hep-ph/0212404}.

\bibitem{Aarts:2001yn}
Gert Aarts and J{\"u}rgen Berges, \emph{Classical aspects of quantum fields far
  from equilibrium}, Phys. Rev. Lett. \textbf{88} (2002) 041603,
\eprint{hep-ph/0107129}.

\bibitem{Aarts:2003bk}
Gert Aarts and Jose~M. Mart{\'i}nez~Resco, \emph{Transport coefficients from
  the 2PI effective action}, Phys. Rev. \textbf{D68} (2003) 085009,
\eprint{hep-ph/0303216}.

\bibitem{Berges:2004hn}
J.~Berges, Sz. Bors{\'a}nyi, U.~Reinosa, and J.~Serreau, \emph{Renormalized
  thermodynamics from the 2PI effective action}, Phys. Rev. \textbf{D71} (2005)
  105004,
\eprint{hep-ph/0409123}.

\bibitem{Schwinger:1960qe}
Julian~S. Schwinger, \emph{Brownian motion of a quantum oscillator}, J. Math.
  Phys.
\textbf{2} (1961) 407.

\bibitem{Bakshi:1962dv}
Pradip~M. Bakshi and Kalyana~T. Mahanthappa, \emph{Expectation value formalism
  in quantum field theory. 1}, J. Math. Phys.
\textbf{4} (1963) 1.

\bibitem{Bakshi:1963bn}
Pradip~M. Bakshi and Kalyana~T. Mahanthappa, \emph{Expectation value formalism
  in quantum field theory. 2}, J. Math. Phys.
\textbf{4} (1963) 12.

\bibitem{Keldysh:1964ud}
L.~V. Keldysh, \emph{Diagram technique for nonequilibrium processes}, Sov.
  Phys. JETP
\textbf{20} (1965) 1018.

\bibitem{Berges:2002wt}
J{\"u}rgen Berges and Markus~M. M{\"u}ller, \emph{Nonequilibrium quantum fields
  with large fluctuations}, in \emph{Progress in Nonequilibrium Green's
  Functions 2} (edited by M. Bonitz and D. Semkat, World Scientific Publ.,
  Singapore, 2003), 367, \eprint{hep-ph/0209026}

\bibitem{Juchem:2004cs}
S.~Juchem, W.~Cassing, and C.~Greiner, \emph{Nonequilibrium quantum-field
  dynamics and off-shell transport for $\phi^4$-theory in 2+1 dimensions},
  Nucl. Phys. \textbf{A743} (2004) 92,
\eprint{nucl-th/0401046}.

\bibitem{Berges:2005vj}
J{\"u}rgen Berges and Szabolcs Bors{\'a}nyi, \emph{Nonequilibrium quantum
  fields from first principles}  (2005),
\eprint{hep-th/0512010}.

\bibitem{Berges:2005md}
J{\"u}rgen Berges and Szabolcs Bors{\'a}nyi, \emph{Range of validity of
  transport equations}  (2005),
\eprint{hep-ph/0512155}.

\bibitem{Berges:2005ai}
J{\"u}rgen Berges, Szabolcs Bors{\'a}nyi, and Christof Wetterich,
  \emph{Isotropization far from equilibrium}  (2005),
\eprint{hep-ph/0505182}.

\bibitem{Dolgov:1997mb}
A.~D. Dolgov, S.~H. Hansen, and D.~V. Semikoz, \emph{Non-equilibrium
  corrections to the spectra of massless neutrinos in the early universe},
  Nucl. Phys. \textbf{B503} (1997) 426,
\eprint{hep-ph/9703315}.

\bibitem{Berges:2004yj}
J{\"u}rgen Berges, \emph{Introduction to nonequilibrium quantum field theory},
  AIP Conf. Proc. \textbf{739} (2005) 3,
\eprint{hep-ph/0409233}.

\bibitem{montvayMunster1994a}
Istv{\'a}n Montvay and Gernot M{\"u}nster, \emph{Quantum fields on a lattice}
  (Cambridge University Press, 1994)

\bibitem{fftw}
Matteo Frigo and Steven~G. Johnson, \emph{FFTW Reference Manual} (Version 3.1,
  2006)

\bibitem{Arrizabalaga:2005tf}
Alejandro Arrizabalaga, Jan Smit, and Anders Tranberg, \emph{Equilibration in
  $\varphi^4$ theory in 3+1 dimensions}, Phys. Rev. \textbf{D72} (2005) 025014,
\eprint{hep-ph/0503287}.

\bibitem{gsl}
Mark Galassi et~al., \emph{GNU Scientific Library Reference Manual} (Version
  1.5, 2004). See also \cite{pressEtAl1988a}.

\bibitem{vanHees:2001ik}
Hendrik van Hees and J{\"o}rn Knoll, \emph{Renormalization in self-consistent
  approximations schemes at finite temperature. I: Theory}, Phys. Rev.
  \textbf{D65} (2002) 025010,
\eprint{hep-ph/0107200}.

\bibitem{vanHees:2001pf}
Hendrik van Hees and J{\"o}rn Knoll, \emph{Renormalization of self-consistent
  approximation schemes. II: Applications to the sunset diagram}, Phys. Rev.
  \textbf{D65} (2002) 105005,
\eprint{hep-ph/0111193}.

\bibitem{vanHees:2002bv}
Hendrik van Hees and J{\"o}rn Knoll, \emph{Renormalization in self-consistent
  approximation schemes at finite temperature. III: Global symmetries}, Phys.
  Rev. \textbf{D66} (2002) 025028,
\eprint{hep-ph/0203008}.

\bibitem{Blaizot:2003an}
Jean-Paul Blaizot, Edmond Iancu, and Urko Reinosa, \emph{Renormalization of
  phi-derivable approximations in scalar field theories}, Nucl. Phys.
  \textbf{A736} (2004) 149,
\eprint{hep-ph/0312085}.

\bibitem{Berges:2005hc}
J{\"u}rgen Berges, Szabolcs Bors{\'a}nyi, Urko Reinosa, and Julien Serreau,
  \emph{Nonperturbative renormalization for 2PI effective action techniques},
  Annals Phys. \textbf{320} (2005) 344,
\eprint{hep-ph/0503240}.

\bibitem{pressEtAl1988a}
William~H. Press et~al., \emph{Numerical Recipes in C} (Cambridge University
  Press, 1988)

\end{thebibliography}
